\documentclass[12pt]{article}
\usepackage{array,amsmath,amssymb,verbatim,bbm}
\usepackage{graphicx}
\usepackage{bm}
\csname @addtoreset\endcsname{equation}{section}

\newcommand{\ket}[1]{\left| #1\right\rangle}               

\newcommand{\braket}[2]{\langle #1|\, #2 \rangle}

\newcommand{\fbraket}[2]{\langle #1  |\! #2 \rangle}
\newcommand{\bra}[1]{\left\langle #1\right|}
\newcommand{\G}{\mathcal{G}}
\newcommand{\s}{\mathcal{S}}
\newcommand{\F}{\mathcal{F}}
\newcommand{\V}{\mathcal{V}}

\begin{document}

{\hbox to\hsize{September 2005 \hfill
{Bicocca--FT--05--22}}}

\begin{center}
\vglue .06in
{\Large \bf Improved Hartree--Fock resummations\\
and spontaneous symmetry breaking}
\\[.45in]
Andrea Sartirana\footnote{andrea.sartirana@mib.infn.it}
~and~
Claudio Destri\footnote{claudio.destri@mib.infn.it}\\
{\it Dipartimento di Fisica dell'Universit\`a degli studi di
Milano-Bicocca,\\ 
and INFN, Sezione di Milano, piazza della Scienza 3, I-20126 Milano,
Italy}\\[.8in]

{\bf ABSTRACT}\\[.0015in]
\end{center}

The standard Hartree--Fock approximation of the $O(N)-$invariant $\phi^4$
model suffers from serious renormalization problems. In addition, when the
symmetry is spontaneously broken, another shortcoming appears in
relation to the Goldstone bosons: they fail to be massless in the
intermediate states. In this work, within the framework of
out--of--equilibrium Quantum Field Theory, we propose a class of systematic
improvements of the Hartree--Fock resummation which overcomes all the
above mentioned difficulties while ensuring also exact
Renormalization--Group invariance to one loop.


\section{Introduction, summary and outlook}
\label{sec:intr-summ-outl}

The research in out--of--equilibrium dynamics of quantum fields has
received, in recent years, a great impulse by cosmology as well as by
particle and condensed matter physics. Indeed, a first--principle
theoretical treatment of many important phenomena, such as the reheating of
the universe after inflation or the thermalization of the quark gluon
plasma in the ultra--relativistic heavy--ion colliders (RHIC, LHC), is
required for a good qualitative and quantitative understanding of the late
time and strongly coupled evolution of quantum field systems.

In particular, this necessity has encouraged the study of nonperturbative
approaches to Quantum Field Theory (QFT) that could provide all--orders
partial resummations of Feynman diagrams
\cite{Berges:intro,Manfredini:2000sk}. In fact standard perturbation theory
does not yield satisfactory results, except for very short times, when
non--equilibrium conditions are involved.

Mean--field approximations such as Leading--order large--N expansion
\cite{Cooper:largeN,Boyanovsky:1994me,Baacke:largeN,Destri:largeN} and
Hartree, or Hartree--Fock (HF) variational method
\cite{Amelino-Camelia:1992nc,Destri:1999he,Michalski,Matsui} are the
simplest and most studied \cite{Berges:2000ur,Habib:irrev,
  Boyanovsky:meanfield,Wetterich:meanfield,Smit:meanfield,Baacke:2001zt}
resummation schemes. Their main features are well known: they do provide a
backreaction term on the evolution of quantum fluctuations that stabilize
dynamics after parametric amplifications or spinodal instabilities, but
they fail to reproduce an important property of late time dynamics such as
thermalization.

A more powerful tool is the 2PI (or 2PPI) effective action
\cite{Calzetta:1986cq,Cornwall:1974vz} whose expansions at two (or more)
loops order or at next--to--leading order in $1/N$ provide resummations
that go beyond mean field approaches
\cite{Dominici,Berges:NLO,Cooper:2005vw,Baacke:BeyondHF} and yield, indeed,
approximate numerical thermalization at strong coupling.

More formal aspects of resummed approximations, such as their
renormalization properties, have been studied as well
\cite{Lenaghan:1999si,Pi:1987df}. Recently a systematic method has been put
forward \cite{Cooper:2004rs,Jakovac,Blaizot,vanHees} that removes
divergences in the $\Phi$--derivable approximations by applying a BPHZ
subtraction procedure to diagrams with resummed propagators.

In \cite{Destri:2005qm} we considered the simple HF approximation of the
$O(N)$ $\phi^4$ model in the unbroken symmetry phase. It is known that it
cannot be consistently renormalized by the usual renormalization of bare
coupling and mass \cite{Destri:1999he,Lenaghan:1999si,Pi:1987df}. We showed
that this non-renormalizability is due to the absence of leading
logarithmically divergent contributions coming from diagrams which do not
have a ``daisy'' topology and therefore are not present in the standard HF
resummation. The inclusion of these contribution together with suitably
chosen finite parts led us to the definition of a renormalized and
Renormalization--Group invariant version of HF equations.

In the present paper we further develop this analysis, making it more
systematic and transparent, and then apply it to the subtler case of
spontaneously broken $O(N)$ symmetry. In this case, together with the
shortcomings as far as pure renormalizability is concerned, the HF
approximation shows an unphysical non--gapless behaviour. Namely, the
effective resummed propagators of the Goldstone bosons, as defined by
the HF equation of motion, are not massless. This is a well known
problem which has been cured by defining suitably modified
approximations \cite{Baym:1977qb,vanHees:2002bv}. In this paper we
define a systematic improvement of the standard HF approach which
ensures gaplessness together with renormalizability and
RG-invariance. In practice we include all leading logarithmically
divergent contributions needed to render finite the resummed
propagator masses (which are $\log{\Lambda}$ dependent in standard HF)
while choosing the finite parts in such a way to have massless
transverse degrees of freedom.

More in detail, we proceed as follows. First, by analysing the diagrammatic
resummation performed by the HF approximation, we identify the missing
leading logarithmically divergent contributions that cause the
non-renormalizability. As will be apparent below, this analysis closely
parallels that in \cite{Destri:2005qm} since, after all, the UV behavior of
the theory is the same regardless of the breaking of the $O(N)$ symmetry.
Then we construct our modified HF approximation, including the missing
contributions, while choosing the corresponding finite parts by means of
the following recipe
\begin{itemize}
\item[-] determine the general features of the HF approach that must be
  shared by the modified approximation, such as mean--field structure,
  $O(N)$ Ward identities, leading--log structure and more, and parameterize
  the class of resummations having such properties.
\item[-] identify amid this family of approximations the one(s) with
  suitable properties of renormalizability, RG--invariance,
  gaplessness and infrared finiteness.
\end{itemize}
One important difference from \cite{Destri:2005qm} is that, when the $O(N)$
symmetry is spontaneously broken, the above procedure does not lead to a
unique result except for the $N=1$ case ($\mathbb{Z}_2$ symmetry).  When
$N>1$ we find a whole class of approximations, corresponding to different
choices of finite parts, that share the required characteristics. Like in
\cite{Destri:2005qm} the improved equations have, by construction, a mean
field structure but, unlike the standard ones, are nonlocal in space and
time.

In sec.~\ref{sec:generalities} we review some general concepts. We
introduce the CTP formulation of out--of--equilibrium problems and define
the HF approximation as resummation of bubble diagrams recalling the
corresponding, well known, equations of motion [see
eqs.~(\ref{eq:HF-eom-G}) and eqs.~(\ref{eq:HF-eom-mf})]. Some important
general features of the mean field approximations are pointed out.

In sec.~\ref{sec:n=1-case} we study, as an illuminating example, the case
of spontaneously broken $\mathbb{Z}_2$ symmetry.
Subsec.~\ref{sec:analys-hf-appr} is dedicated to a general analysis of the
approximation. In subsec.~\ref{sec:renorm-rg-invar} we apply the standard
renormalization procedure and point out its shortcomings. In
subsec.~\ref{sec:defin-modif-hf} we define our modified HF approximation
which is renormalizable and RG--invariant.  As we already said in this case
we are led to a unique result [see eqs.~(\ref{eq:mHF-pardef-n=1}) and
eqs.~(\ref{eq:mHF-eom-mf})]. Further remarks on the freedom of choosing
various initial conditions without spoiling renormalization and
RG--invariance are made in subsec.~\ref{sec:other-initial-states}.

In sec.~\ref{sec:on-case} we generalize to $N>1$, that is to a continuous
$O(N)$ spontaneous symmetry breaking, by first analysing some general
features [subsec.~\ref{sec:analys-hf-appr-1}] of the standard HF
approximation and its renormalization properties
[subsec.~\ref{sec:renorm-rg-invar-1}] and then defining our modified
approximation by the general recipe given above
[subsec.~\ref{sec:defin-modif-hf-1}]. The final results will be a set of
constraints that should be satisfied by renormalizable and RG--invariant
Hartree--Fock--like approximations together with two examples [see
eqs.~(\ref{eq:mHF-ex-1}) and eqs.~(\ref{eq:mHF-ex-2})].

There are several possible developments along the lines of this work. First
of all it would be interesting to study other properties of the different
improved approximations that we have constructed in the $N>1$ case and,
hopefully, identify further constraints that would allow to single out a
unique result. Secondly, a numerical study of the modified HF field
equations found in this paper and in \cite{Destri:2005qm} would be needed
to investigate how the space-time nonlocalities affect the time evolution
as compared to the standard HF approximation, which is known to fail even
qualitatively at late times. Another challenging task is the extension of
our approach to the full two--loop 2PI effective action, through the
inclusion of the nonlocal sunset diagram which is absent by definition in
any mean--field approximation. In fact one should expect that, even if such
an inclusion allows to recover renormalizability as compared to the
conventional HF approximation, the two--loop 2PI self-consistent equations
still lack RG invariance with the two--loop beta function, since the 2PI
effective action does not contain all the diagrams which contribute to the
next--to--leading ultraviolet divergences.

\section{Generalities}
\label{sec:generalities}

To study the real--time dynamics from a given set of initial
conditions we need to evaluate matrix elements of the form
\begin{equation*}
   \bra{\Psi(t_0)}| \mathcal{O} ( t_1,t_2,\dots) \ket{\Psi(t_0)}
\end{equation*}   
where $\ket{\Psi}$ is a generic state prepared at a initial time $t_0$ and
$\mathcal{O}$ is some operator depending on the field evaluated at times
$t_i > t_0$. Out--of--equilibrium QFT provides the general setup for the
calculation such matrix elements, as well as more general expectation
values in statistical mixtures of pure states such as $\ket{\Psi}$.  In
this section we briefly review some generalities about non--equilibrium
QFT, define the Hartree--Fock approximation and point out some properties
that will be useful to derive the forthcoming results. In doing this, we
restrict to the case of interest for this paper: scalar field theory in
$3+1$ dimensions with quartic interaction and spontaneously broken $O(N)$
symmetry.

\vskip15pt
The field variables are $\varphi_i ( \bm{x} )$ where $\bm{x}=
(x_1,x_2,x_3)$ are space coordinates and $i=1,\dots,N $ is the $O(N)$
index. Classical dynamics is defined by an action, functional of
trajectories $\varphi_i ( x )$, with $x= ( \bm{x}, t )$, in the form
\begin{equation*}
  S[\varphi]= \int d^4 x \left\{ \frac{1}{2} \partial_\mu 
  \varphi_i ( x ) \partial^\mu \varphi_i ( x ) - \frac{m^2}{2} 
  \varphi_i ( x ) \varphi_i ( x ) - \frac{\lambda}{4!} [\varphi_i ( x ) 
  \varphi_i ( x ) ]^2 \right\}
\end{equation*}  
where $m^2$ and $\lambda$ are the squared mass (negative in the case
of spontaneously broken symmetry ) and the coupling constant
respectively.

\vskip15pt 
In QFT, the general approach to non--equilibrium dynamics was developed by
Keldish and Schwinger and is known as closed time path (CTP) formalism.  It
allows to use standard functional methods
(see~\cite{Calzetta:1986cq,Cornwall:1974vz,Cooper:1995zs,Chou:1984es}).  by
introducing path integrals on a time path going from $t=0$ to $t=+\infty$
and back. Field integration variables are then doubled and subdivided into
$(+)-$components, for the path integral forward in time, and
$(-)-$components for the backward piece.  Given an initial state defined by
the functional $\Psi$ of the field configurations $\varphi (\bm x)$, one
writes down the functional integral
\begin{equation}
  \label{eq:CTP-path-int}
  e^{i\,\mathcal{W}[j_+,j_-]} = \int \mathcal{D}\varphi_+ 
  \mathcal{D} \varphi_- \Psi[\varphi_+]  \overline{\Psi} [ \varphi_-]
  \,e^{\,i S[\varphi_+]-i S[\varphi_-]+ i \fbraket{j_+}{\,\varphi_+}
    -i\fbraket{j_-}{\,\varphi_-}}
\end{equation}
where we have used the short-hand notation
\begin{equation*}
  \langle a  | M  | b  \rangle \equiv \int d^3 x\, 
  d^3 y~ a_j( \bm{x})  M_{jk} ( \bm{x},\bm{y})~b_k (\bm{y}) 
\end{equation*}
which will be useful later on. Integration in eq.~(\ref{eq:CTP-path-int})
is on trajectories from $t=0$ to $t=+ \infty$ (with the condition
$\varphi_+ = \varphi_-$ at $t= +\infty$) and $\varphi_\pm$ in the wave
functional is the $t=0$ section of $\varphi_\pm$.  Notice that, in the
action functional $S$ that enters in the path integral in
eq.~(\ref{eq:CTP-path-int}), the constant parameters $\lambda$, $m^2$ are
to be substituted with the bare (cut-off dependent) ones $\lambda_0$,
$m^2_0$. In fact the theory should be thought as regularized with an $UV$
cut-off $\Lambda$ and then renormalized to remove the divergent dependence
on $\Lambda$. In the general setting, field renormalization should be
included as well.  Here we ignore it since, as we will see, it is absent in
the HF approximation of the scalar theory.

By construction $\mathcal{W}[j_+,j_-]$ is
the generating functional of connected Green functions
\begin{equation*}
  \begin{split}
    G_{+\dots+-\dots-}(x_1,\dots,x_n,y_1,\dots,y_n) &\equiv
    \frac{ (-i)^{n+m}\,\delta^{n+m} \mathcal{W}}
    {\delta j_+(x_1)\dots \delta j_+(x_n)\delta j_-(y_1)
      \dots \delta j_-(y_m)} \Bigg|_{\substack{j_+ = 0 \\j_- = 0}} \\[2mm] 
    &= -i\bra{\Psi} \overline{\mathcal{T}} \{\varphi(y_1)\dots \varphi(y_m)\} 
    \mathcal{T}\{\varphi (x_1)\dots \varphi(x_n) \}\ket{\Psi}_{\rm conn}
  \end{split}
\end{equation*}
where $\mathcal{T}$ and $\overline{\mathcal{T}}$ define time ordered and
inverse ordered products, respectively, and internal indices have been
omitted for ease of notation. The effective action $\Gamma_{\rm
  1PI}[\phi_+,\phi_-]$, which is the generator of 1PI vertex functions, is
the Legendre transform of $\mathcal{W}[j_+,j_-]$ from the currents
$j_{\pm}$ to the fields $\phi_{\pm}$. The equation of motion for the
background field $\phi(x)=\bra{\Psi}\varphi(x)\ket{\Psi}$ then reads
\begin{equation}
  \label{eq:CTP-bk-eom}
  \frac{\delta \Gamma_{\rm 1PI}}{\delta \phi_+ (x)}
  \Bigg|_{\phi_+=\,\phi_-=\phi}=0
\end{equation}   
Notice that the functional $\mathcal{W}$ as well as the 1PI effective
action parametrically depend on the initial state $\Psi$. In our present
discussion we consider an initial wave functional having the following
Gaussian form
\begin{equation*}
  \Psi[\varphi]=
  \mathcal{N} \exp\left\{i \braket{{\dot \phi}(0)}{\varphi-\phi(0)} - 
    \bra{\varphi - \phi(0)}\left[\tfrac14 \G^{-1}+ i\s \right] 
  \ket{\varphi-\phi(0)}\right\}
\end{equation*}
whose free parameters are the $t=0$ background field $\phi(\bm{x},0)$, the
$t=0$ background momentum $\dot \phi(\bm{x},0) $, the real symmetric
positive kernel $\G_{ij} ( \bm{x},\bm{y})$ and the real symmetric kernel
$\s_{ij} ( \bm{x},\bm{y})$. We can see that $\Gamma_{\rm 1PI}$ now depends
parametrically only on the kernels $\G$ and $\s$, while the $t=0$
background fields $\phi(0)$ and $\dot \phi(0)$ enter instead as initial
conditions for the equation eq.~(\ref{eq:CTP-bk-eom}).

\vskip15pt
The perturbative diagrammatic expansion in the CTP formalism proceeds as in
vacuum QFT. The bare propagators are
\begin{equation*}
  \begin{split}
    G^{(b)}_{++} (x,y) &\equiv G^{(b)}_{F}(x,y) = -i \bra{\Psi}\mathcal{T}
    \varphi( x )\varphi( y )\ket{\Psi}_{\rm conn} |_{\rm free} \\
    G^{(b)}_{--} (x,y) &\equiv G^{(b)}_{\bar F}(x,y) = -i
    \bra{\Psi}\overline{\mathcal{T}} \varphi( x )\varphi( y
    )\ket{\Psi}_{\rm conn} |_{\rm free} \\ G^{(b)}_{+-}(x,y) &=
    G^{(b)}_{-+}(y,x) = -i\bra{\Psi}\varphi( y )\varphi( x )
    \ket{\Psi}_{\rm conn} |_{\rm free}
  \end{split}
\end{equation*} 
Here the notation $|_{\rm free}$ indicates that, in accordance to the tree
level of the theory in the broken symmetry phase, the Heisenberg
expectation values have to be calculated for a free scalar field theory
with mass matrix $m^{2}_{b,ij}=\tfrac13\,\lambda_0\,v_i\,v_j$, where $v_i$
is the nonvanishing vacuum expectation value. The bare vertices are
\begin{equation*}
  \centering
  \includegraphics[width=.5\textwidth]{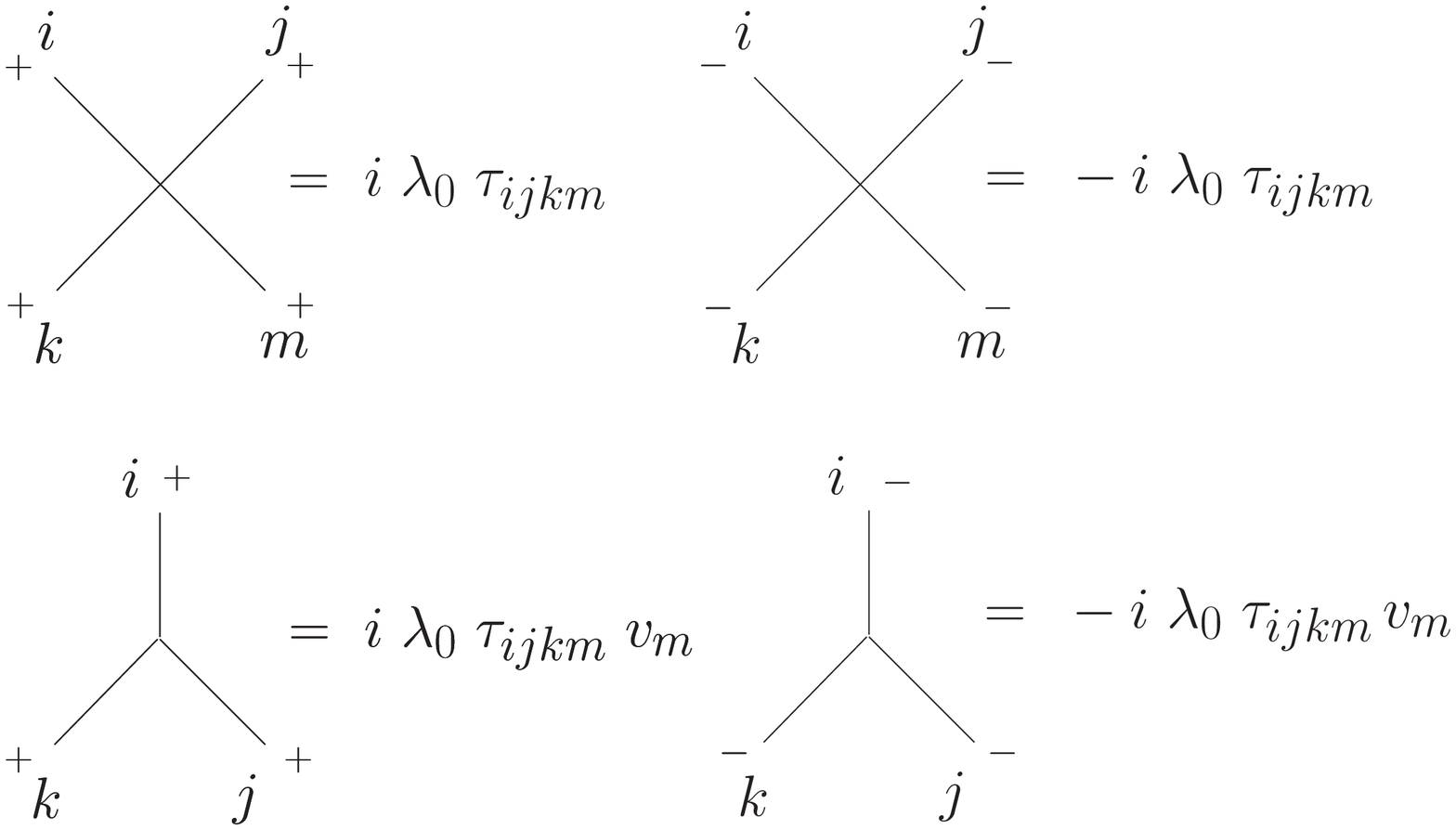}
\end{equation*}
However, in order to obtain sensible results in a generic
out--of--equilibrium contexts or even at equilibrium with nonzero
temperature, it is known that calculations should go beyond plain
perturbation and perform (partial) resummations to all orders in the
coupling constant. A very successful resummation method is provided by the
introduction of the 2PI effective action
(see~\cite{Calzetta:1986cq,Cornwall:1974vz}). It is defined as the double
Legendre transform of the $\mathcal{W}$ generating functional with respect
to the usual current one--point $j_{\pm}$ and to the two--points current
$K_{\alpha \beta}(x,y)$ coupled through the term
\begin{equation*}
  \frac{1}{2} \int d^4 x \int d^4 y ~K_{\alpha \beta}( x, y )
  \varphi_{\alpha}(x) \varphi_{\beta} ( x )
\end{equation*}
where $\alpha, \beta = \pm $. $\Gamma_{\rm 2PI}$ is a functional of the
classical fields $\phi_{\alpha}$ and of the propagators $G_{\alpha \beta}$.
It yields two equations of motions
\begin{equation}
\label{eq:CTP-2PI-eom}
  \frac{\delta \Gamma_{\rm 2PI}}{\delta \phi_{\alpha} ( x )}\Big|_*= 0 \quad,
  \qquad \frac{\delta \Gamma_{\rm 2PI}}{\delta G_{\alpha \beta}( x,
    y)}\Big|_*= 0 
\end{equation}
Here the notation $|_*$ indicates that, by their physical meaning, the
$(\pm)$--component fields and propagators have to satisfy, on the
solutions of motion, the following relations 
\begin{equation*}
  \begin{split}
    \phi_-(x) &= \phi_-(x)=\phi(x) \\ G_{F}(x,y)&= G_{+-}(y,x) \theta
    (x_0 - y_0) + G_{+-}(x,y) \theta (y_0 - x_0) \\ G_{\bar F}(x,y)&=
    G_{+-}(x,y) \theta (x_0 - y_0) + G_{+-}(y,x) \theta (y_0 - x_0)
  \end{split}
\end{equation*}
Hence the system eq.~(\ref{eq:CTP-2PI-eom}) reduces to two coupled
equations for $\phi$ and $G_{+-}$ only. Moreover, any initial Gaussian
state may be absorbed in the $t=0$ term for the $j_\alpha$ and $K_{\alpha
  \beta}$ currents, so that, by the double Legendre transform the initial
Gaussian state disappears from the effective action, but fixes the initial
conditions on $\phi$ and $G_{+-}$. The role of $\phi(0)$ and $\dot\phi(0)$
is immediate, while for the kernels we have
\begin{equation*}
  \begin{split}
    G_{+-}(x,y)|_{x_0=y_0=0}&= \G (\bm x,\bm y)\\
    \frac{\partial}{\partial y_0} G_{+-}(x,y)|_{x_0= y_0=0} &= 2 i
    \,[\G \s]({\bm x},{\bm y})+ \tfrac12 \delta^3 (\bm{x}- \bm{y})
  \end{split}
\end{equation*} 
Given $\Gamma_{\rm 2PI}$ at a certain perturbative loop order, if we
solve the second equation in eqs.~(\ref{eq:CTP-2PI-eom}) for a generic
$\phi$ and substitute the result $G[ \phi ]$ into the first one we
obtain the background equations of motion corresponding to a resummed
diagrammatic approximation of the 1PI effective action $\Gamma_{\rm
1PI}$.

In the present scalar theory $\Gamma_{\rm 2PI}$ has the general form
\begin{equation*}
  \Gamma_{\rm 2PI} \left[ \phi, G \right]= S[\phi] + \frac{i}{2} \mathrm{Tr} 
  \left[ \log G \right] + \frac{i}{2} \mathrm{Tr} \left[ G_0^{-1} G \right] + 
  \Gamma_2 \left[ \phi, G \right]
\end{equation*}
where $S$ is the complete classical action of the double time path (i.e.
$S= S_+ - S_-$). Traces are taken over all indices $i$, $\alpha$ and $x$.
$G_0^{-1}$ is the second derivative of the action in a $\phi$ background,
$\Gamma_2$ is the sum of all vacuum 2PI diagrams with $G$ propagators and
vertices defined by the classical action in a $\phi$ background. To two
loops level the diagrams contributing to the $\Gamma_2$ are the ``8'' and
``sunset'' diagrams 
\begin{equation*}
  \centering
  \includegraphics[scale=0.16]{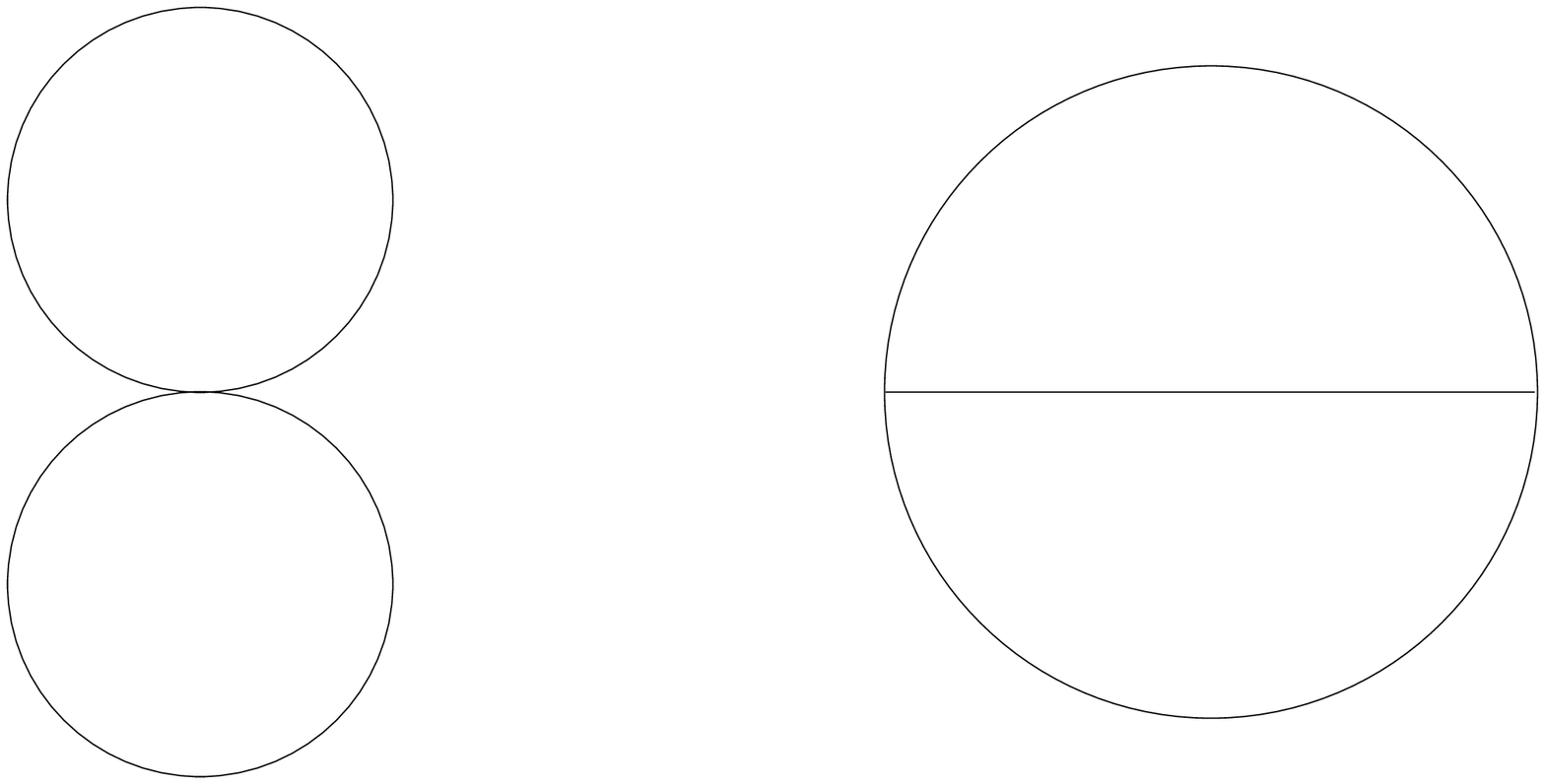}
\end{equation*}
Now we can introduce the Hartree--Fock approximation as obtained
considering only the first contribution (i.e. the ``8'' graph) to
$\Gamma_2$.
\begin{equation*}
   \Gamma_2= \tfrac{i}{8} \lambda_0 \left[ G^2_{F} ( x,x) - G^2_{\bar F}
   ( x,x) \right] 
\end{equation*}
Notice that ``8'' is the only 2PI diagram made of ``product'' of loops
corresponding to a mean field contribution to the mass. In the 1PI
framework this corresponds to a resummation of all vacuum 1PI diagrams
with daisy and superdaisy topologies of the form
\begin{equation*}
  \centering
  \includegraphics[scale=0.5]{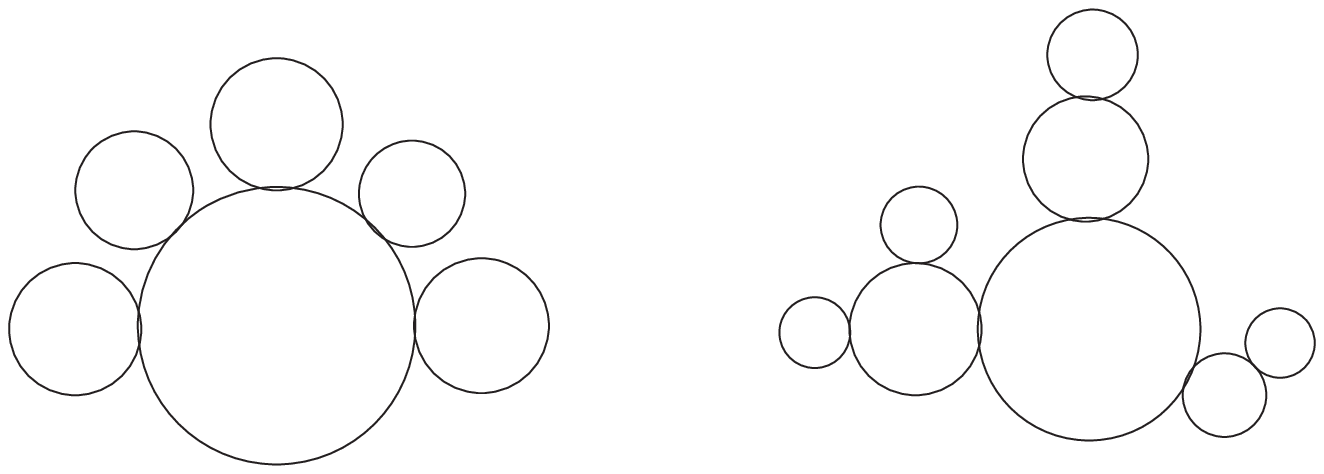}
\end{equation*}
By explicitly using this form of $\Gamma_2$ in
eqs.~(\ref{eq:CTP-2PI-eom}), setting $G(x,y)= \tfrac{i}{2} [
G_{-+}(x,y) + G_{-+}(y,x)]$ and observing that the antisymmetric
combination decouples, one obtains
\begin{equation}
  \label{eq:HF-eom-G} 
  \begin{split}
    &\left\{ \left[ \Box + m_0^2 +\tfrac16\lambda_0
      \,\phi_k(x)\phi_k(x) \right] \delta_{ij} + \tfrac12
    \lambda_0\, \tau_{ijkm} G_{km} (x, x) \right\} \phi_j(x)=0
    \\[2mm] & \left\{ \left( \Box + m_0^2 \right) \delta_{ij} +
    \tfrac12 \lambda_0 \,\tau_{ijkm} \big[ \phi_k(x) \phi_m(x) +
      G_{km}(x,x) \big]\,\right\} G_{ij}(x,y) =0
  \end{split}
\end{equation}
where the tensor $\tau$ is defined as  
\begin{equation}
  \label{eq:tau-tens-def}
  \tau_{ijkm} = \tfrac13( \delta_{ij} \delta_{km} + \delta_{ik}
  \delta_{jm} + \delta_{im} \delta_{jk} )
\end{equation}
It is convenient to introduce an equivalent formulation of the
eqs.~(\ref{eq:HF-eom-G}) in terms of mode functions $u_{\bm{k}\,a}$
($\bm{k}$ is the wave vector and $a$ the $O(N)$ polarization), that will be
the one used throughout the paper. For simplicity let us suppose that the
initial ($t=0$) kernels are translationally invariant (while the background
field and momentum may be point dependent). We can then write
\begin{equation*}
  \G_{ij} ( \bm{x}, \bm{y}) = \int \frac{d^3 k }{(2\pi)^3}\,
  \tilde{\G}_{ij} ( \bm{k} )~ e^{i \bm{k} \cdot \bm{x}} \quad,\qquad
  \s_{ij} ( \bm{x}, \bm{y}) = \int \frac{d^3 k }{(2 \pi)^3}\,
  \tilde{\s}_{ij} ( \bm{k} )~ e^{i \bm{k} \cdot \bm{x}}
\end{equation*}
Next, we introduce the $t=0$ mode functions by 
\begin{equation*}
  u_{\bm{k},ai}(\bm{x},0) = [\tilde{\G}( \bm{k})^{1/2}]_{ai} \,e^{i
    \bm{k} \cdot \bm{x}} \quad,\qquad \dot{u}_{\bm{k},ai}(\bm{x},0) =
  \left[- \tfrac{i}2 \tilde{\G}(\bm{k})^{-1} + 2\, \tilde
    \s(\bm{k})\right]_{ij} u_{\bm{k},aj} (\bm{x}, 0)
\end{equation*}
Then, one can easily verify that, by the identification
\begin{equation*}
  G_{ij}(x,y) = \mathrm{Re} \int \frac{d^3 k }{(2\pi)^3}~
  u_{\bm{k},ai}(x)\,\overline{u}_{\bm{k},aj}(y)
\end{equation*}   
eqs.~(\ref{eq:HF-eom-G}) are equivalent to the following equations of
motion
\begin{equation} 
  \label{eq:HF-eom-mf}
  \begin{split}
    &\left\{ [ \Box + m_0^2 +\tfrac16\lambda_0 \,\phi_k(x)\phi_k(x) ]
      \delta_{ij} + \tfrac12 \lambda_0\, \tau_{ijkm} \int \frac{d^3 p
      }{(2\pi)^3}~ u_{\bm{p},bm}(x) \,\overline{u}_{\bm{p},bn}(x)\right\}
      \phi_j(x)=0 \\[2mm] & \left\{ ( \Box + m_0^2 ) \delta_{ij} +
      \tfrac12\lambda_0 \tau_{ijmn} \left[ \phi_m(x) \phi_n(x) + \int
      \frac{d^3 p }{(2\pi)^3}~ u_{\bm{p},bm}(x)
      \,\overline{u}_{\bm{p},bn}(x) \right]\right\} u_{\bm{k},aj} (x) =0
  \end{split}
\end{equation}
These are the HF equations of motions that will be used in the rest 
of the paper.

\vskip15pt 
We conclude this section by introducing a different representation of the
CTP formalism, known as the physical representation
(see~\cite{Chou:1984es}) which allows us to derive some results useful
below. We introduce the field redefinitions (omitting again the internal
indices to simplify notation)
\begin{equation*}
  \phi_{\Delta} = \phi_+ - \phi_- \quad,\qquad
  \phi_{c} = \frac{1}{2} ( \phi_{+} + \phi_{-} )
\end{equation*}
and write the 1PI effective action as a functional of these new fields,
$\Gamma_{\rm 1PI}= \Gamma[\phi_\Delta,\phi_c]$. By calculating vertex
functions, one then finds
\begin{equation}
  \label{eq:PR-prop-1}
  \frac{\delta^{n} \Gamma}{\delta \phi_c ( x_1 ) \dots 
    \delta \phi_c ( x_n )} \Bigg|_{\phi_\Delta = 0}= 0
\end{equation}
and 
\begin{equation}
  \label{eq:PR-prop-2}
  \frac{\delta^{n+m} \Gamma}{\delta \phi_c ( x_1 ) \dots \delta 
    \phi_c ( x_n )\delta \phi_\Delta ( y_1 ) \dots \delta \phi_\Delta 
    ( y_m )} \Bigg|_{\phi_\Delta = 0}= 0
\end{equation}
if the time component of anyone of the $x$'s is larger than the time
component of all the $y$'s. Let us also remark that, by the
definition of CTP generating functional, all time coordinates in the
vertex functions are supposed to be positive so we can set, as well,
these functions to be zero for any negative time. Using
eq.~(\ref{eq:PR-prop-1}) together with eq.~(\ref{eq:CTP-bk-eom}) one
can write the equations of motions in the form
\begin{equation}
  \label{eq:PR-bk-eom} 
  \frac{\delta \Gamma}{\delta \phi_{\Delta}}
  \Bigg|_{\phi_\Delta=0,~\phi_c=\phi}=0
\end{equation}
Notice that $2n$--legs vertex functions with one $\phi_\Delta$ leg and
$2n-1$ $\phi_c$ legs are the only ones contributing to these equations
of motion.  Then eq.~(\ref{eq:PR-prop-2}) guarantees that all the
terms nonlocal in time in eq.~(\ref{eq:PR-bk-eom}) do satisfy
causality.

Perturbative calculations by diagrammatic expansion in the physical
representation are based on the bare propagators
\begin{equation}
  \label{eq:PR-bare-props}
  \begin{split}
    G^{(b)}_{c \Delta} (x,y) &\equiv G^{(b)}_{A}(x,y) = -i \theta ( y_0 - x_0 )
    \bra{\Psi}[\phi( x ),\phi( y )]\ket{\Psi}  |_{free} \\
    G^{(b)}_{\Delta c} (x,y) &\equiv G^{(b)}_{R}(x,y) = -i \theta ( x_0 - y_0 )
    \bra{\Psi}[\phi( x ),\phi( y )]\ket{\Psi}  |_{free} \\
    G^{(b)}_{\Delta \Delta} (x,y) &= -2i \bra{\Psi}
    \{\phi( y ),\phi( x )\}\ket{\Psi}_{\rm conn}  |_{free} \;,\quad
    G^{(b)}_{c c} (x,y) = 0
  \end{split}
\end{equation} 
The bare retarded and advanced Green functions $G_A^{(b)}$ and $G_R^{(b)}$
do not depend on the initial state and are translational invariant.  The
bare correlation function $G^{(b)}=\tfrac{i}{4}G^{(b)}_{\Delta \Delta}$,
instead, does depend on $\ket{\Psi}$. The vertices are
\begin{equation*}
  \centering
  \includegraphics[width=0.5\textwidth]{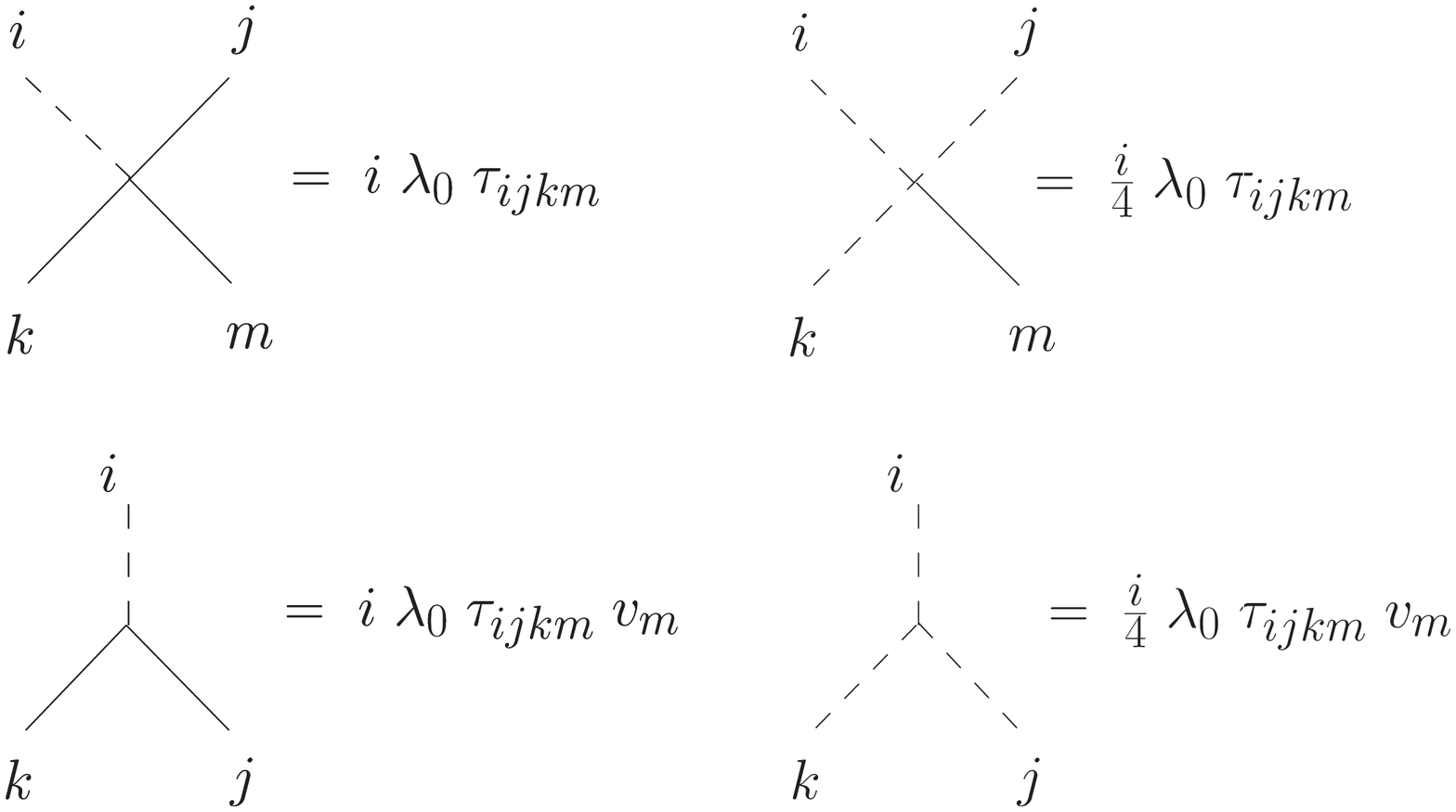}
\end{equation*}
where solid lines represent $\phi_c$ legs while dotted lines represent
$\phi_\Delta$ legs.

\vskip15pt
As stated above, the HF approximation consists in the resummation of
diagrams with daisy and superdaisy topologies. By considering this
diagrammatic resummation it is easy to verify that the corresponding
effective action has the following general structure
\begin{equation*}
  \Gamma_{\rm HF}[\phi_\Delta, \phi_c]= -\bra{\phi_\Delta}\Box\ket{\phi_c} - 
  \F[ \xi, \chi, \eta]
\end{equation*}
where $\F$ is a functional of the following composite matrix fields 
\begin{equation*}
  \xi_{ij} ( x ) = \phi_{c,i}( x )\phi_{c,j}( x )\,,\quad \chi_{ij} (
  x ) = \phi_{c,i}( x )\phi_{\Delta,j}( x )\,,\quad \eta_{ij} ( x ) =
  \phi_{\Delta,i}( x )\phi_{\Delta,j}( x )
\end{equation*}
We now introduce, for ease of notation, the new object
\begin{equation*}
  \F'[\xi]_{ij} ( x ) = \frac{1}{2}
  \frac{\delta \F}{\delta \chi_{ij}(x)} \Bigg|_{\chi=\eta=0}
\end{equation*}
which is a functional of
$\xi_{ij}(x)=\phi_{c,i}(x)\phi_{c,j}(x)=\phi_i(x)\phi_j(x)$ only. Then
the equation of motion in the HF approximation takes the form
\begin{equation*}
  \frac{\delta \Gamma}{\delta \phi_{\Delta}}
  \Bigg|_{\phi_\Delta=0,~\phi_c=\phi}=\{ \Box \delta_{ij} + 2
  \,\F'_{ij}[\xi]\big\} \phi_j=0
\end{equation*}
We will consider this as a general form for mean--field--type
background field equations and it will be that base for our definition
of a modified HF approximation.

\section{The case $N=1$}\label{sec:n=1-case}

We begin by considering a single scalar field theory with spontaneously
broken $\mathbb{Z}_2$ symmetry. The study of this simpler case provides
valuable insight into the general features of the diagrammatic resummation
performed by the HF approximation. In particular, this allows to understand
the origin of the HF shortcomings with respect to renormalizability and
RG--invariance and to determine the general recipe for the definition of a
modified renormalizable and RG-invariant mean field approximation. Many of
the results of this section will hold true also in the more general case of
a theory with $N$ scalar fields.

\subsection{Analysis of the HF approximation}\label{sec:analys-hf-appr} 
For the N=1 case the Hartree-Fock equations of motion [see
eqs.~(\ref{eq:HF-eom-mf})] reduce to
\begin{equation}
  \label{eq:hf-eom-n=1} 
  \begin{split} 
    &\left\{ \Box + m_0^2 +\tfrac16\lambda_0 \,\xi (x) +
    \tfrac12 \lambda_0\, \int_{p^2<\Lambda}\frac{d^3 p }{(2\pi)^3}~ 
    |u_{\bm{p}}(x)|^2\right\}
    \phi(x)=0 \\[2mm] 
    & \left\{ \Box + m_0^2 + \tfrac12 \lambda_0 \,\xi (x) + \tfrac12
    \lambda_0 \,\int_{p^2<\Lambda}\frac{d^3 p }{(2\pi)^3}~
    |u_{\bm{p}}(x)|^2 \,\right\} u_{\bm{k}}(x) =0
  \end{split}
\end{equation}
where we recall that $\xi (x)= \phi^2(x)$. Notice that, so far, these
are still the equations of the regularized theory, written in terms of bare
parameters ($\lambda_0$, $m^2_0$) and explicitly dependent on the
sharp cut-off $\Lambda$.

The phase with spontaneously broken symmetry, which is the subject of this
paper, is defined by assuming the existence of a static homogeneous vacuum
solution of eqs.~(\ref{eq:hf-eom-n=1}) with nonzero background field
$\phi(x) = v$. The corresponding mode functions, that will be named
$u^{({\rm vac})}$, have the following plane wave form
\begin{equation}
  \label{eq:vac-sol-n=1}
  u^{({\rm vac})}_{{\bm k}}(x) = \frac{1}{\sqrt{2\, \omega_{\bm k}}}\,
  e^{i(\bm{k}\cdot\bm{x}-\omega_{\bm k} t)} \quad, \qquad
  \omega^2_{\bm k} = {\bm k}^2 + m^2
\end{equation}
where $m^2=\tfrac13\lambda_0 v^2$ and the vacuum expectation value $v$
of the background should satisfy the gap equation
\begin{equation}
  \label{eq:gap-eq-n=1}
  0 = m^2_0 + \tfrac16\,\lambda_0\,v^2 + \tfrac12 \,\lambda_0
  \int_{p^2<\Lambda} \frac{d^3 p}{(2 \pi)^3} \,\frac{1}{2\, \omega_{\bm k}}
\end{equation} 
The values of bare parameters for which this equation admits a
nonzero solution $v$ are those corresponding to 
spontaneously broken symmetry and are those we are considering here.

We now introduce the mean field $\V$ according to
\begin{equation}
  \label{eq:V-def-n=1}
  \V(x) = m_0^2 +\tfrac16\,\lambda_0 v^2 + \tfrac12\,\lambda_0 \,
  \Delta \xi (x) + \tfrac12\,\lambda_0 \,\int_{p^2<\Lambda} \frac{d^3
    p }{(2\pi)^3}~ |u_{\bm{p}}(x)|^2 
\end{equation}
where $\Delta \xi = \xi - v^2$. Notice that $\V=0$ on the static solution
$\phi = v$ and $u=u^{({\rm vac})}$.  This definition allows us to rewrite
eqs.~(\ref{eq:hf-eom-n=1}) in the form
\begin{equation}
  \label{eq:eom-comp-n=1} 
  \begin{split} 
    &\left\{ \Box - \tfrac13\lambda_0 \,\Delta \xi (x) + \V(x) \right\}
    \phi(x)=0 \\[2mm]
    & \left\{ \Box + m^2 + \V(x) \right\} u_{{\bm k}}(x) =0    
  \end{split}
\end{equation}
Recalling the definition of the background field equation in
terms of the 1PI effective action 
\begin{equation}
  \label{eq:gen-bkeq-n=1}
  \frac{\delta \Gamma_{\rm 1PI}}{\delta \phi_\Delta}
  \Big|_{\substack{\phi_\Delta = 0 \\\phi_c = \phi}} =\big(\Box + 2
  \,\F'[\xi]\big) \phi=0
\end{equation}
and comparing it with the first of eqs.~(\ref{eq:eom-comp-n=1}), we read
out
\begin{equation}
  \label{eq:FV-rel-n=1}
  \F'[\xi]= \tfrac12\,\V-\tfrac16\,\lambda_0\,\Delta\xi
\end{equation}
Notice that $\V$ is regarded here as a functional of $\xi=\phi^2$. The
implicit dependence on $\xi$ is determined by solving the second equation
in~(\ref{eq:eom-comp-n=1}) with a generic background.  To obtain a
self--consistent equation for the functional $\V[\xi]$ from the HF
equations of motion some preliminary definitions are required.
\begin{itemize}
\item
We introduce the free mode functions $u_{{\bm k}}^{(0)}$ defined as
solutions of the free equation with mass $m^2 =
\tfrac13\,\lambda_0\,v^2$ and with the same initial conditions of the
exact mode functions, that is
\begin{equation*}
  ( \Box + m^2 )u_{{\bm k}}^{(0)} ( x ) = 0\quad,\qquad u_{{\bm
      k}}^{(0)}(\bm{x}, 0) = u_{{\bm k}}(\bm{x}, 0)\quad,\qquad \dot
  u_{{\bm k}}^{(0)}(\bm{x}, 0) = \dot u_{{\bm k}}(\bm{x}, 0)
\end{equation*}
\item 
In terms of these we introduce the free correlation
\begin{equation}
  \label{eq:free-corr-n=1}
  G^{(0)}(x,y)=  \mathrm{Re} \int_{|\bm p|<\Lambda}
  \frac{d^3 p }{(2 \pi)^3}\,u_{\bm{p}}^{(0)}(x)\,
  \overline{u}_{\bm{p}}^{(0)}(y)
\end{equation}
\item
Finally we define the free retarded and advanced Green functions as
\begin{equation*}
  G_{R}^{(0)} ( x - y ) =G_{A}^{(0)} ( y - x ) = \int \frac{d^4 p
  }{(2 \pi)^4} \frac{1}{ p^2 - m^2 + i \epsilon p_0 }~ e^{-i p ( x -
    y )}
\end{equation*}
\end{itemize}
By using the above definitions we can cast the equation of motion for
the mode functions into a convenient integral form
\begin{equation*}
  u_{{\bm k}}(x) = u_{{\bm k}}^{(0)}(x) + \int
  d^4y\,G^{(0)}_{R} (x-y)\,\V (y) \, u_{{\bm k}}(y)
\end{equation*}
In this equation, and everywhere else from now on, all field ($\phi$,
mode functions, $\F'$, ${\cal V}$, etc...) are to be thought as
defined only for positive times (initial conditions are at the limit
point $t=0^+$) and all time integrations are restricted to positive
values, as appropriate in an initial value problem.  Moreover it is
convenient to introduce a more compact operator notation with implicit
space--time integration, that is
\begin{equation}
  \label{eq:mfeq-intform-n=1}
  u_{{\bm k}} = u_{{\bm k}}^{(0)} + G_R^{(0)}\hat{\V}\, u_{{\bm k}}
\end{equation}
where the caret $\,\hat{}\,$ turns a vector of the
functional space into a multiplication operator
\begin{equation*}
  \hat{}\,: \, v(x) \rightarrow \hat{v}(x,y) = v(x)\, \delta^{(4)}(x-y) 
\end{equation*} 
>From now on we shall use this notation throughout the paper.

Eq.~(\ref{eq:mfeq-intform-n=1}) can be formally solved for $u_{\bm k}$
\begin{equation*}
  u_{{\bm k}} = [{\bf 1}- G_R^{(0)}\hat{{\cal V}}]^{-1}u_{{\bm k}}^{(0)} 
\end{equation*}
where ${\bf 1}$ stands for the space time delta function
$\delta^{(4)}(x-y)$. Then the cutoffed correlation 
\begin{equation*}
  G (x,y) = \mathrm{Re} \int_{|\bm p|<\Lambda} \frac{d^3 p }{(2 \pi)^3}
  \, u_{\bm{p}}(x)\, \overline{u}_{\bm{p}}(y)
\end{equation*}
can be written as 
\begin{equation}
  \label{eq:corr-V-n=1}
  G = G[{\cal V}] = [{\bf 1}-G_R^{(0)}\hat{\cal V}]^{-1}\,G^{(0)}\,
  [{\bf 1}-\hat{\cal V} G_A^{(0)}]^{-1}
\end{equation}
in terms of $\V$, of the free retarded and advanced Green functions
and of the free correlation function. In conclusion, by
renaming the correlation at coincident points as
\begin{equation}
  \label{eq:I-def-n=1}
  I[{\cal V}](x) = G[{\cal V}](x,x)
\end{equation}
and substituting into the definition of $\V$ in eq.~(\ref{eq:V-def-n=1}) we
obtain the sought self--consistent equation 
\begin{equation}
  \label{eq:scon-eq-n=1}
  \V = m_0^2 +\tfrac16\,\lambda_0 v^2 + \tfrac12 \lambda_0 \, \big(
    \Delta \xi  + I[{\cal V}] \big)
\end{equation}
Notice that eq.~(\ref{eq:scon-eq-n=1}) depends parametrically on the
initial kernels (i.e. the mode functions initial conditions) through
the explicit form of $G^{(0)}$. Now, before going any further in the
discussion, we fix a particular choice for these kernels by
considering the HF vacuum as initial state for the quantum
fluctuations. That is to say that we start from equilibrium initial
conditions for the mode functions
\begin{equation*}
  u_{{\bm k}}(\bm{x}, 0) = u_{{\bm k}}^{({\rm vac})}(\bm{x},
  0)\quad,\qquad \dot u_{{\bm k}}(\bm{x}, 0) = \dot
  u_{{\bm k}}^{({\rm vac})}(\bm{x}, 0)
\end{equation*}
which corresponds to the following choice of the initial kernels
\begin{equation*}
  \tilde{\G} ({\bm k} ) = \frac{1}{2 \omega_{\bm k}} \; \quad,
  \qquad \tilde{\s} ( {\bm k} )=0
\end{equation*}
By this choice some simplifying properties follow
\begin{itemize}
\item
The free mode functions coincide with the vacuum mode functions
$u^{({\rm vac})}$ at every time.
\item
By the self--consistent equation we have $\V=0$ at the point $\xi (x)
= v^2$
\item
The free correlation function defined in eq.~(\ref{eq:free-corr-n=1})
turns to be translationally invariant in spacetime. Explicitly
\begin{equation*}
  G^{(0)}(x-x') = \int_{\bm{k^2}< \Lambda^2} \frac{d^3 k}{(2
    \pi)^3}~\frac{1}{2 \omega_{\bm k}} \cos[\omega_{\bm k}(t-t') - {\bm
      k}\cdot({\bm x}- {\bm x}')]
\end{equation*}
\end{itemize}
\noindent
Finally let us stress that, in spite of this choice, we are still
considering an out--of--equilibrium problem since we allow for generic
initial conditions for the background field.  The use of different initial
kernels will be discussed later.

\vskip 10pt 
At this point some observations on the integral term $I$ defined by
eq.~(\ref{eq:I-def-n=1}) and eq.~(\ref{eq:corr-V-n=1}) are in order.  

\vskip10pt
\noindent
{\bf 1}.~~$I$ depends on the free retarded (and advanced) Green function
and on the free correlation function. These propagators have the same form
of their bare correspondents [see eq.~(\ref{eq:PR-bare-props})] but their
mass is $m^2=\tfrac13 \lambda_0 v^2$ rather than $m^2_b=m^2_0+\tfrac12
\lambda_0 v^2$. By using the gap eq.~(\ref{eq:gap-eq-n=1}), the relation
between the two sets of propagators reads
\begin{equation}
  \label{eq:barfre-rel-n=1}
  \begin{split}
    &G_R^{(0)}= G_R^{(b)} + \tfrac12
    \lambda_0\,G_R^{(0)}\,\hat{I}^{(1)}\,G_R^{(b)}\\[2mm]
    &G^{(0)}=[{\bf
    1}-\tfrac12\,\lambda_0\,G^{(0)}_R\,\hat{I}^{(1)}]\,G^{(b)}\, [{\bf
    1}-\tfrac12\,\lambda_0\,\hat{I}^{(1)}\,G^{(0)}_A]
  \end{split}
\end{equation}
where $I^{(1)}$ is the spacetime constant tadpole of the $G^{(0)}$ propagator
\begin{equation*}
  I^{(1)}=G^{(0)}(x,x)
\end{equation*}
By recursively solving eq.~(\ref{eq:barfre-rel-n=1}) for the free
propagators we can see that their definition corresponds to the
resummation of all tadpole corrections. As we will see they play the
role of internal dressed propagators in the diagrammatic resummation
performed by the Hartree-Fock approximation.  
\vskip10pt
\noindent
{\bf 2}.~~By expanding $I$ in powers of $\V$ we can see that the term
proportional to the $n$-th power contains loop with $n+1$
free propagators. In particular the expansion up to the linear term
reads
\begin{equation*}
  I[\V](x) =\left[G^{(0)} +G_R^{(0)}\,\hat{\cal V}\,G^{(0)} + G^{(0)}\,
    \hat{\cal V} \, G_A^{(0)}\right](x,x) + \dots
\end{equation*}
which, for later convenience, can be rewritten in the compact form
\begin{equation*}
  I[\V] = I^{(1)} + I^{(2)}\,\V + \dots
\end{equation*}
in terms of the tadpole $I^{(1)}$ and of the two propagators loop
\begin{equation*}
  I^{(2)}(x-y) = + 2 \, G^{(0)}_{R}(x,y)\, G^{(0)}(x,y)
\end{equation*}
\vskip10pt
\noindent {\bf 3}.~~One can easily realize that $I^{(1)}$ diverges as
$\Lambda^2$ and $\log \Lambda$, $I^{(2)}$ diverges as $\log\Lambda$ and the
loops with more than two propagators are convergent. It is therefore useful
to introduce a new functional $J$ containing only the convergent part of
$I$ according to
\begin{equation}
  \label{eq:J-def-n=1}
  J[\V] = I[\V]- I^{(1)} - I^{(2)}\,\V
\end{equation}
\vskip10pt
\noindent
The previous observations lead us to rewrite the self--consistent
eq.~(\ref{eq:scon-eq-n=1}) in a particular ``quasi-renormalized'' form,
more suitable to analyze the diagrammatic resummation of the effective
action and determine its divergent graphs and subgraphs structure.  The
``quasi-renormalized'' form is obtained by expanding the self--consistent
eq.~(\ref{eq:scon-eq-n=1}) around the point $\xi (x) = v^2$ ($\V=0$),
explicitly solving the linear terms and writing a self--consistent equation
for the higher $\Delta \xi$ powers dependence of $\V$.

First we substitute eq.~(\ref{eq:J-def-n=1}) into
eq.~(\ref{eq:scon-eq-n=1}), applying the gap eq.~(\ref{eq:gap-eq-n=1}) in
order to simplify the constant terms. Then, by solving the terms
proportional to $\V$ we obtain
\begin{equation}
\label{eq:theta-def-n=1}
  \V =\tfrac12 \,\theta \, \big[ \Delta \xi + J[\V] \big] \quad ,
  \qquad \theta = \lambda_0 \, \left[ {\bf 1} -
  \tfrac12\,\lambda_0 \, I^{(2)} \right]^{-1}
\end{equation}
or, equivalently
\begin{equation}
  \label{eq:qr-Veq-n=1}
  \V= \tfrac12 \,\theta\, \Delta \xi + \Delta \V \quad , \qquad \Delta
  \V =\tfrac12 \,\theta \, J[\tfrac12\,\theta\,\Delta \xi+ \Delta \V]
\end{equation}
In conclusion we rewrite eq.~(\ref{eq:FV-rel-n=1}) in terms of
$\Delta\V$ and of the explicit linear term as
\begin{equation}
  \label{eq:VF-qrrel-n=1}
  \F'=\tfrac12\,\Omega^{(2)}\,\Delta \xi + \tfrac12\,\Delta \V \quad,
  \qquad \Omega^{(2)} = \tfrac12 \, \theta - \tfrac13\,
  \lambda_0\,{\bf 1}
\end{equation}
We can now use eqs.~(\ref{eq:qr-Veq-n=1}) and~(\ref{eq:VF-qrrel-n=1}) to
calculate the vertex functions relevant for the background equation of
motion. In the physical representation of the CTP formalism these functions
have the form
\begin{equation}
  \label{eq:gamma-CPT-n=1}
  \Gamma^{(n)}(x_1,\dots,x_n)=
  \frac{\delta\Gamma_{1PI}[\phi_\delta,\phi_c]}{\delta
    \phi_\Delta(x_1)\delta\phi_c(x_2)\dots \delta
    \phi_c(x_n)}\Big|_{\substack{\phi_\Delta = 0 \\\phi_c = v}}
\end{equation}
with one $\phi_\Delta$ leg at the point $x_1$ and $n-1$ $\phi_c$ legs at
the points $x_i$ ($i>1$).  

First of all, from the general form of the background equation
(\ref{eq:gen-bkeq-n=1}) we can see that the vertex
functions~(\ref{eq:gamma-CPT-n=1}) are built from the functional
derivatives of $\F'$ w.r.t. $\xi$ at the point $\xi=v^2$. To ease notation
we rename such variations by
\begin{equation*}
  \Omega^{(k)}(x_1,\dots,x_{k})= 2\,\frac{\delta\F'[\xi](x_1)}{\delta
    \xi(x_2)\dots \delta \xi(x_k)}\Big|_{\xi=v^2}
\end{equation*}
Some examples of the expressions of the vertex functions in terms of
the $\Omega^{(k)}$ are
\begin{equation}
  \label{eq:Gamma-Omega-n=1}
  \begin{split} 
    &\Gamma^{(1)}(x)=\Omega^{(1)}(x)\,v \quad,\qquad
    \Gamma^{(2)}(x_1,x_2)=[\Box + \Omega^{(1)}(x_1)]\,\delta_{1,2} +
    2\,\Omega^{(2)}_{12}\,v^2\\[2mm]
    &\Gamma^{(3)}(x_1,x_2,x_3)=2\,\Omega^{(2)}_{12}
    \,\delta_{12}\,v+2\, \Omega^{(2)}_{13}\,\delta_{23}\,v+
    2\,\Omega^{(2)}_{12}\, \delta_{13}\,v+4\,
    \Omega^{(3)}_{123}\,v^3\\[2mm]
    &\Gamma^{(4)}(x_1,x_2,x_3,x_4)=2\{\,\Omega^{(2)}_{13}\,
    \delta_{12}\, \delta_{34}+ \Omega^{(2)}_{13}\, \delta_{14}\,
    \delta_{23}+\Omega^{(2)}_{12}\, \delta_{13}\, \delta_{24}\}+
    8\,\Omega^{(4)}_{1234}\,v^4+\dots\\[2mm]
    &\dots+4\{\,\Omega^{(3)}_{134}\,\delta_{12}
    +\Omega^{(3)}_{124}\,\delta_{13}+\Omega^{(3)}_{123}\,\delta_{14}
    +\Omega^{(3)}_{124}\,\delta_{23}+
    +\Omega^{(3)}_{123}\,\delta_{34}+\Omega^{(3)}_{123}\,\delta_{24}
    \}\,v^2
  \end{split}
\end{equation}
where we have introduced the shorthand notations
\begin{equation*}
  \delta_{ik}=\delta^{(4)}(x_i-x_k) \quad,\qquad
  \Omega^{(2)}_{i\dots k}=\Omega^{(2)}(x_i,\dots,x_k)
\end{equation*}
>From eq.~(\ref{eq:VF-qrrel-n=1}) we can see that $\Omega^{(1)}=0$. So
that $\Gamma^{(1)}=0$, which is just the statement that $v$ is the
vacuum solution of the background equation. Moreover, by
eq.~(\ref{eq:VF-qrrel-n=1}) we have
\begin{equation}
  \label{eq:Om2-HF-n=1}
  \Omega^{(2)}= \tfrac12\,\theta-\tfrac13\,\lambda_0\,{\bf 1}=
  \tfrac16 \, \lambda_0\,{\bf 1} - \tfrac14\,\lambda_0^2\,
  I^{(2)}\,[{\bf 1}-\tfrac12\,\lambda_0\,I^{(2)}]^{-1}
\end{equation}
Expanding the above expression in powers of $\lambda_0$ we see that
$\Omega^{(2)}$ is the sum of a classical term $\tfrac16\,\lambda_0$ plus the
resummation of all the chains of $I^{(2)}$ loops. 

The $\Omega^{(k)}$ with $k>2$ are obtained by using
eq.~(\ref{eq:VF-qrrel-n=1}) and the self--consistent equation for
$\Delta\V$ [eq.~(\ref{eq:qr-Veq-n=1})]. One can easily see that they are
built up with loops having three or more propagators and are therefore
finite in the limit $\Lambda\to\infty$. The loops are
attached to each other and to the external legs by the effective vertex
$\theta$ [see eq.~(\ref{eq:theta-def-n=1})]. Some graphical examples are
\begin{equation*}
  \centering
  \includegraphics[scale=0.4]{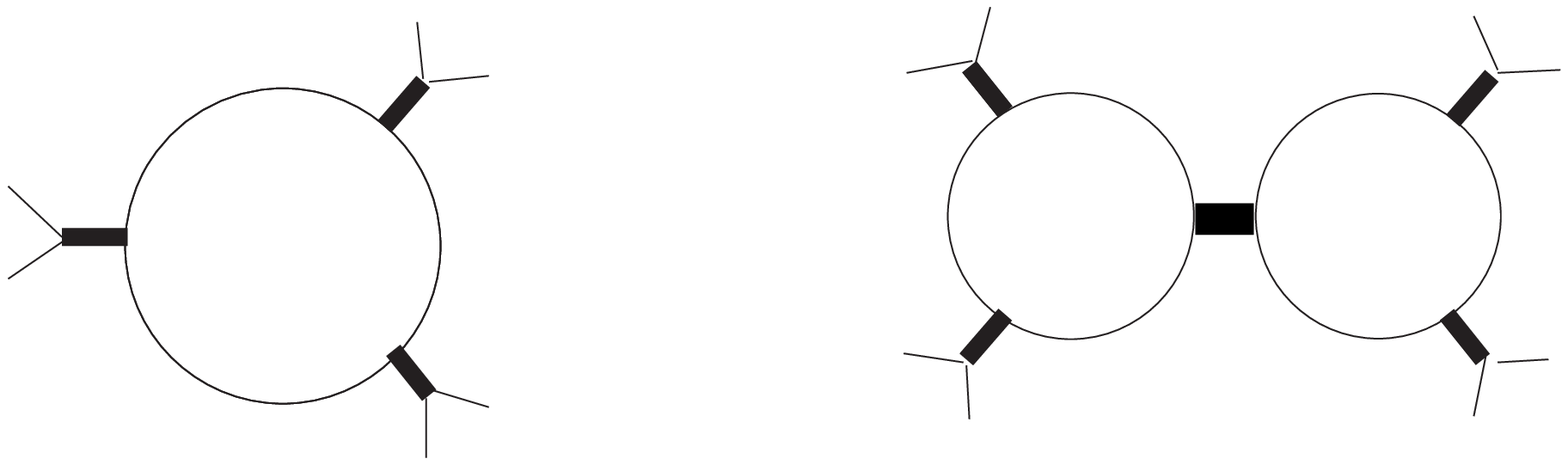}
\end{equation*}
The vertex $\theta$, by its explicit definition in
eq.~(\ref{eq:theta-def-n=1}), is the sum of the classical term $\lambda_0$
plus chains of $I^{(2)}$ integrals
\begin{equation}
  \label{eq:theta-HF-n=1}
  \theta= \lambda_0\,{\bf 1}- \tfrac12\,\lambda_0^2\,
  I^{(2)}\,[{\bf 1}-\tfrac12\,\lambda_0\,I^{(2)}]^{-1}
\end{equation}

\subsection{Renormalization}
\label{sec:renorm-rg-invar}
Let us now apply the standard renormalization procedure to the HF
approximation of the effective action. As we will see it fails to
render all the vertex functions finite.

The first renormalization condition is simply the request that the v.e.v. $v$
has a physical cutoff--independent value. Then the gap eq.~(\ref{eq:gap-eq-n=1})
must be regarded as defining the bare mass parameter $m_0^2$ as a function of
$\lambda_0$, $v$ and the cut-off $\Lambda$
\begin{equation}
  \label{eq:mass-renorm-N=1}
  m^2_0(\lambda_0,v,\Lambda) =- \tfrac16\,\lambda_0\,v^2 - \tfrac12
  \,\lambda_0 \int_{p^2<\Lambda} \frac{d^3 p}{(2 \pi)^3} \,\frac{1}{2\,
    \omega_{\bm k}}
\end{equation} 
Substituting the above function for the constant $m_0^2$ in the
equation of motion turns to be enough to remove all the $\Lambda^2$
dependence in the vertex functions.

The second renormalization condition follows by fixing the value
of the quartic coupling at some fixed energy scale. In order to do
this we introduce the Fourier transform of the vertex
functions [see eq.~(\ref{eq:gamma-CPT-n=1})], namely 
\begin{equation*}
  \tilde \Gamma^{(m)}(p_1,\dots,p_{m-1})=\int
  \Big[\prod_{n=2}^m d^4 x_n\Big]
  \Gamma^{(m)}(x_1,\dots,x_m)\, e^{i\sum_m p_n\cdot
    (x_1-x_n)}
\end{equation*}
where translational invariance, which is a consequence of our choice of
initial conditions, has been assumed.
For later convenience we define also the Fourier transform of $\Omega^{(k)}$
\begin{equation*}
  \tilde \Omega^{(k)}(p_1,\dots,p_{k-1})=\int
  \Big[\prod_{n=2}^{k} d^4 x_n\Big]
  \Omega^{(k)}(x_1,\dots,x_{k})\, e^{i\sum_n p_n\cdot
    (x_1-x_n)}
\end{equation*}
We recall that, as appropriate to a causal initial value problem, all
Fourier transforms are analytic in the upper complex $p_0$--halfplane.

Now, in the present out--of--equilibrium context, we can define the symmetric
point at the scale $s$ by setting the momenta entering the $\phi_c$
legs ($p_2$, $p_3$, $p_4=-p_1-p_2-p_3$) to the same, purely
spatial, value
\begin{equation}
  \label{eq:simm-point}
  -\tfrac13\, p_1= p_2 = p_3 =p_4=\tfrac12\,q_s \;,\quad q_s \equiv (0,\bm
  q)\;,\quad |\bm q|=s
\end{equation}
with an arbitrary direction ${\hat{\bm q}}$. The usual coupling
renormalization condition is obtained by evaluating $\Gamma^{(4)}$
at this point and requiring it to be equal to the renormalized
coupling constant $\lambda$
\begin{equation}
  \label{eq:recon-cougen-n=1}
  \lambda=\tilde \Gamma^{(4)}(-\tfrac32\,q_s,\tfrac12\,q_s,\tfrac12\,q_s)
\end{equation}
Now, recalling the general expression of $\Gamma^{(4)}$ in terms of
the $\Omega^{(k)}$ in eq.~(\ref{eq:Gamma-Omega-n=1}) , and performing
the Fourier transform, we can rewrite eq.~(\ref{eq:recon-cougen-n=1})
as
\begin{equation}
  \label{eq:recon-coustd-n=1}
  \begin{split}
    \lambda=&\,6\,\tilde \Omega^{(2)}(q_s)+ 4\{\, \tilde
    \Omega^{(3)}(\tfrac12 q_s,\tfrac12 q_s) + 2\,\tilde
    \Omega^{(3)}(-\tfrac32 q_s,\tfrac12 q_s) + \tilde
    \Omega^{(3)}(-q_s,\tfrac12 q_s)\\ &+ \tilde \Omega^{(3)}(\tfrac12
    q_s,-q_s)\}\,v^2+8\, \tilde \Omega^{(4)}(-\tfrac32 q_s,\tfrac12
    q_s,\tfrac12 q_s)v^4
  \end{split}
\end{equation}
However, in our case we prefer to slightly change this standard procedure
in order to avoid lengthy calculations and obtain a better comparison with
the unbroken symmetry case. We therefore substitute
eq.~(\ref{eq:recon-coustd-n=1}) simply with
\begin{equation}
  \label{eq:recon-coumod-n=1}
  \lambda=6\,\tilde \Omega^{(2)}(q_s)
\end{equation}
We have omitted in this way the contributions to the quartic coupling
originating from the superficially convergent $\Omega^{(3)}$ and
$\Omega^{(4)}$.  Hence the change from eq.~(\ref{eq:recon-coustd-n=1}) to
eq.~(\ref{eq:recon-coumod-n=1}) is to be regarded as a legitimate finite
part redefinition of the coupling constant.

Now, by the HF expression for $\Omega^{(2)}$, eq.~(\ref{eq:Om2-HF-n=1}),
the renormalized coupling constant explicitly reads
\begin{equation}
  \label{eq:HF-btorrel-n=1}
  \lambda=\lambda_0\, \frac{1+\lambda_0\,\tilde I^{(2)}(q_s)}{1 -
    \tfrac12\,\lambda_0 \, \tilde I^{(2)}(q_s)}
\end{equation}
This has to be compared with the corresponding formula of unbroken symmetry
case in ref.~\cite{Destri:2005qm}. Notice that in ref.~\cite{Destri:2005qm}
we have adopted the opposite sign convention for $I^{(2)}$.
Eq.~(\ref{eq:HF-btorrel-n=1}), once inverted, defines $\lambda_0$ as a
function of $\lambda$, $v$ and (the logarithm of) $\Lambda$.

\vskip15pt 
As already stated above, the renormalization procedure just outlined fails
to define a sensible renormalized theory. In particular we can individuate
the following shortcomings.

\vskip10pt
There is clearly a pathological behavior of the effective quartic coupling
$\lambda$ as a function of the bare parameter $\lambda_0$ at fixed cut-off
$\Lambda$. In fact $\lambda$ has the correct $1-$loop $\lambda_0^2$ term
dictated by perturbation theory, but certainly fails at higher orders,
since is exhibits an unphysical behaviour, growing to a maximum value at
$\lambda_0=\lambda_0^{max}$, then decreasing to zero and to even more
unphysical negative values (returning positive only for very large values
of $\lambda_0$). This implies the breakdown of the HF approximation for
values of $\lambda_0$ greater than $\lambda_0^{max}$ in a theory at fixed
cut-off, as compared, for instance, with the standard
$1-$Loop--Renormalization--Group improved relation which reads
\begin{equation}
  \label{eq:1LRGI-btorrel-n=1}
  \frac{1}{\lambda} = \frac{1}{\lambda_0} + \frac3{16 \pi^2} \,\log
  \Lambda + \dots
\end{equation}
where the dots stand for some suitable choice of the renormalization scale
and of the finite parts. The relation in eq.~(\ref{eq:1LRGI-btorrel-n=1})
is monotonically increasing with $\lambda_0$, at fixed $\Lambda$, to an
asymptotic plateau value $\lambda_{max}$. As a consequence it can be
inverted determining a single bare coupling $\lambda_0$ value for any
$\Lambda$ and $\lambda<\lambda_{max}$.  A full trajectory of $\lambda_0$ as
a function of $\Lambda$, up to the vertical asymptote at the Landau Pole
value $\Lambda_{LP}$, corresponds to a single renormalized theory
describing the same dynamics at momentum scales much smaller than
$\Lambda$. On the other hand, in the Hartree-Fock approximation, the
non--monotonic behaviour of the relation in eq.~(\ref{eq:HF-btorrel-n=1})
spoils this one--to--one correspondence between bare and renormalized
parameters (at fixed cut--off) which holds true only for small coupling.

\vskip10pt
Moreover, we see that imposing a finite value to $\lambda$ at the
chosen scale $s$ fails to render finite the running coupling (i.e. the
generalization of eq.~(\ref{eq:recon-coumod-n=1}) to any value of
momentum)
\begin{equation}
\label{eq:run-coupl-n=1}
  \lambda(p)=6\,\tilde \Omega^{(2)}(p)
\end{equation}
at any momentum $p\neq q_s$ and even at any $q_{s'}$ with $s'\neq s$.

\vskip10pt
For what concerns the higher order terms $\Omega^{(k)}$ with $k>2$ we
can see that a logarithmic dependence on the cut--off persists even
after imposing the renormalization condition in
eq.~(\ref{eq:recon-coumod-n=1}) . That is they do not parametrically
depend solely on $v$ and $\lambda$, but also on $\lambda_0$ and
therefore on $\log\Lambda$. This happens for two specific reasons
\begin{itemize}
\item[1.]The internal propagators $G_R^{(0)}$, $G_A^{(0)}$ and $G^{(0)}$
  introduce an explicit and non removable dependence on $\lambda_0$ due to
  the presence of the resummed mass $m^2=\tfrac13\,\lambda_0\,v^2$. Notice
  that, unlike the unbroken symmetry case, here the resummed contribution
  to the internal propagators differs from those to the external propagator
  defined as the functional inverse of the two--legs vertex function. To
  the latter contribute also diagrams as 
  \begin{equation*}
    \centering
    \includegraphics[scale=0.4]{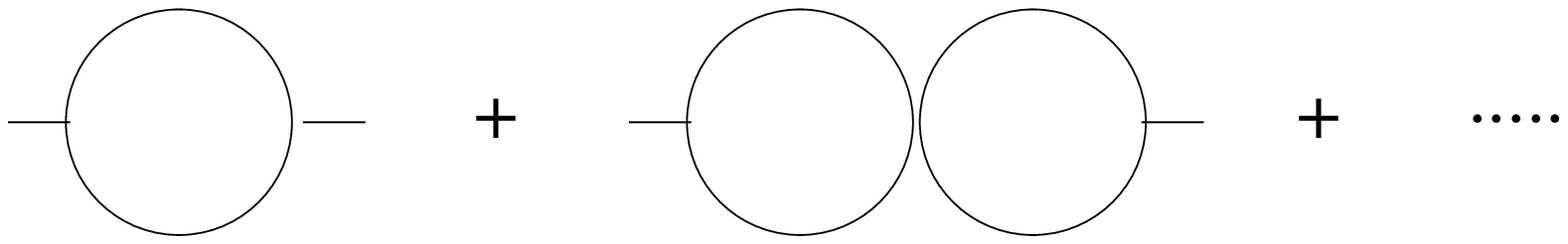}
  \end{equation*}
  as can be seen from its explicit form [see eq.~(\ref{eq:Gamma-Omega-n=1})]
  \begin{equation}
    \label{eq:HF-2legs-n=1} 
    \tilde \Gamma^{(2)}(p)=-p^2+ \tfrac13\,\lambda(p)\,v^2=-p^2+ m^2 -
    \tfrac12\,\lambda_0^2\, \tilde
    I^{(2)}(p)\,[1-\tfrac12\,\lambda_0\,\tilde I^{(2)}(p)]^{-1}\,v^2
  \end{equation} 
  This is finite at momentum $q_s$ while, according with what we said above
  about $\lambda(p)$, logarithmic divergences appear as $p\neq q_s$.
\item[2.]In the HF definition of $\Omega^{(k)}$ with $k>2$ there appears the
  effective vertex $\theta$. By the relation [see
  eq.~(\ref{eq:theta-HF-n=1})]
  \begin{equation*}
    \tilde \theta(p)=\tfrac13\,\lambda(p) + \tfrac23\,\lambda_0
  \end{equation*} 
  we can see that, after imposing the renormalization condition, an
  unresolved $\lambda_0$ dependence persists even when $p=q_s$. For $p\neq
  q_s$ further cut-off dependences appears due the problems concerning
  the renormalized running coupling in eq.~(\ref{eq:run-coupl-n=1}).
\end{itemize} 
We conclude that the the HF approximation cannot be renormalized
by the standard renormalization procedure: there is an unphysical
bare--to--renormalized coupling relation and a plain failure to eliminate
divergences in the subgraphs of the resummation. In the next subsection we
will define a modified HF resummation by explicitly requiring
renormalizability and a $1-$loop--renormalization--group improved relation
between $\lambda$ e $\lambda_0$ [see eq.~(\ref{eq:1LRGI-btorrel-n=1})]. By
comparing the two resummations we will recognize the cause of the
renormalization problems of the Hartree-Fock approximation in the
incomplete resummation of Leading Logarithms of the cut-off.

\subsection{Improved HF approximation}
\label{sec:defin-modif-hf}

Our recipe for the improvement of the HF resummation consists in two
fundamental steps
\begin{itemize}
\item[1.]Fix some features, proper of the HF approximation, that we want to
  maintain throughout the modification since they provide a minimal
  definition of a mean field resummation. The results of
  sec.~\ref{sec:analys-hf-appr} provide a parameterization of the
  class of approximations having such defining features.
  
\item[2.]Require explicitly renormalizability and RG--invariance in order
  to fix the form of the arbitrary parameters.
\end{itemize}
In conclusion, once the result is obtained, we will be able to
establish a diagrammatic interpretation of our modified approximation.
\vskip15 pt
\noindent
The main features of the HF approximation are encoded in
\begin{itemize}
\item The general mean field form of the background equation of motion
\begin{equation}
\label{eq:gen-feat1-n=1}
  \frac{\delta \Gamma_{\rm 1PI}}{\delta \phi_\Delta
   (x)}\Big|_{\substack{\phi_\Delta = 0 \\\phi_c = \phi}} =\big( \Box
   + 2 \,\F'[\xi](x)\big) \phi(x)=0
\end{equation}
\item The self consistent definition of $\F'$ (through the mean field $\V$)  
\begin{equation}
  \label{eq:gen-feat2-n=1}
  \F'=[\tfrac12\, \Omega^{(2)}-\tfrac14 \theta]\,\Delta\xi+
  \tfrac12 \V \quad,\qquad\V=\tfrac12\,\theta\Delta\xi +
  \tfrac12\,\theta\,J[\V]
\end{equation}
where we recall the definition of $J$ by eq.~(\ref{eq:J-def-n=1}) and
eq.~(\ref{eq:I-def-n=1}).
\end{itemize}
As we have seen in sec.~\ref{sec:analys-hf-appr} the relations in
eqs.~(\ref{eq:gen-feat1-n=1}) and~(\ref{eq:gen-feat2-n=1}) provide a
general recipe for building up all vertex functions using
$\Omega^{(2)}$, the effective vertex $\theta$ and the free
propagators. Moreover, the self--consistent
eqs.~(\ref{eq:gen-feat2-n=1}) for $\F'$ correspond to the following
general form of the equation of motion for the mode functions
\begin{equation*}
  \left\{ \Box + m^2 + \V(x) \right\} u_{{\bm k}}(x) =0 \quad,\qquad
  \V = \tfrac12\,\theta\,\{\Delta \xi + I - {\hat I}^{(1)}-
  I^{(2)}\,\V \}
\end{equation*}
with
\begin{equation*}
  \quad I(x)=\int_{p^2<\Lambda} \frac{d^3 p }{(2\pi)^3}~
  |u_{\bm{p}}(x)|^2
\end{equation*}
We now regard $\Omega^{(2)}$, $\theta$ and the effective mass $m^2$ of the
free propagators as tunable parameters defining a class of approximations that
share the same general diagrammatic structure. In other words, we abandon
the HF definitions for these parameters, that read
\begin{equation}
  \label{eq:HF-pardef-n=1}
  m^2=\tfrac13\,\lambda_0\,v^2 \quad,\qquad 
  \theta = \lambda_0 \, \left[{\bf 1} -
    \tfrac12\,\lambda_0 \, I^{(2)} \right]^{-1} \quad,\qquad 
  \Omega^{(2)} =\tfrac12 \, \theta - \tfrac13\,\lambda_0\,{\bf 1}
\end{equation}
and look instead for new definitions that ensure proper renormalizability and
RG--invariance.

We can actually further specify the form of the tunable parameters. By
looking at the HF definitions in eq.~(\ref{eq:HF-pardef-n=1}) of
$\Omega^{(2)}$ and $\theta$ we see that they must have the general leading
log structure
\begin{equation}
\label{eq:genLL-1-n=1}
  \Omega^{(2)}=\lambda_0\,F_1(\lambda_0\,I^{(2)})\qquad,\quad  
  \theta=\lambda_0\,F_2(\lambda_0\,I^{(2)})
\end{equation}
in terms of two functions, $F_1$ and $F_2$, of a single variable (the
evaluation of these functions on the operator $I^{(2)}$ is obvious if we
consider the Fourier transform).  We assume that the same holds true for
$m^2$. This can be understood looking at the explicit form of the two--legs
function in eq.~(\ref{eq:HF-2legs-n=1}). We consider $m^2$ as a function of
$I^{(2)}$ evaluated at zero momentum since we don't want to introduce any
new mass scale dependence, that is 
\begin{equation}
\label{eq:genLL-2-n=1}
  m^2 =\lambda_0\,F_3(\lambda_0\,\tilde I^{(2)}(0))\,v^2
\end{equation}
Notice that changing $m^2$ corresponds to changing the equilibrium
solution according to eq.~(\ref{eq:vac-sol-n=1}).

To conclude we observe that the HF approximation resums correctly the
$1-$loop perturbative order. To maintain this feature we should
require that $\Omega^{(2)}$ matches at tree and $1-$loop order while
$\theta$ and $m^2$ match at tree level. Explicitly we have the
following conditions on the $F$'s
\begin{equation}
  \label{eq:match-cnst-n=1}
  F_1(x)=\tfrac16 + \tfrac14\,x + O(x^2)\;,\quad 
  F_2(x)=1 + O(x)\;,\quad
  F_3(x)=\tfrac13 + O(x)\;,\quad 
\end{equation}
\vskip15pt
\noindent
Now we have to define our modified HF approximation by fixing new explicit
definitions for the functions $F$'s. We obtain renormalizability with the
correct $1-$loop beta function, by assuming the following logarithmic
dependence for the bare coupling $\lambda_0$ 
\begin{equation*}
  \frac{\partial\lambda_0}{\partial \log{\Lambda}} = \frac3{16
    \pi^2}\lambda_0^2 + O(\Lambda^{-1}) 
\end{equation*}
and requiring that the parameters $\Omega^{(2)}$, $\theta$ and $m^2$ do not
depend on $\log{\Lambda}$. Hence, differentiating
eqs.~(\ref{eq:genLL-1-n=1}) and~(\ref{eq:genLL-2-n=1}) with respect to
$\log{\Lambda}$, yields
\begin{equation}
  \label{eq:RG-conds-n=1}
  (1 - \tfrac32\,x)\,F'_I (x)+\tfrac32\,F_I(x)=0\to
  F_I(x)= \frac{A_I}{1-\tfrac32\,x}\qquad I=1,2,3
\end{equation}
where $A_I$ are integration constants and we have used
\begin{equation*}
  \frac{\partial\,I^{(2)} }{\partial \log{\Lambda}} = -\frac1{8
    \pi^2}\,{\bf 1} + O(\Lambda^{-1})
\end{equation*}
Eqs.~(\ref{eq:RG-conds-n=1}) have a unique solution that fulfill the
matching constraints in eqs.~(\ref{eq:match-cnst-n=1}).
\begin{equation*}
  F_2(x)=6\,F_1(x)=3\,F_3(x)=\frac{1}{1- \tfrac32\,x}
\end{equation*}
In terms of the parameters $\Omega^{(2)}$, $\theta$, $m^2$\
\begin{equation}
  \label{eq:mHF-pardef-n=1}
  \theta=6\,\Omega^{(2)}=\lambda_0\,[{\bf
  1}-\tfrac32\,\lambda_0\,I^{(2)}]^{-1}\quad,\qquad m^2= \tfrac13 \,
  \lambda_0\,[1 - \tfrac32 \,\lambda_0\, \tilde I^{(2)}(0)]^{-1}\,v^2
\end{equation}
These are the explicit forms of the parameters in our improved HF
approximation.  
\vskip15pt
\noindent
Now applying the renormalization condition in
eq.~(\ref{eq:recon-coumod-n=1}) with the new definition of
$\Omega^{(2)}$ we obtain the following bare--to--renormalized relation
\begin{equation}
\label{eq:mHF-btorrel-n=1}
  \lambda = \frac{\lambda_0}
  {1 - \tfrac32\,\lambda_0\,\tilde I^{(2)}(q_s)} 
\end{equation}
Comparing this with eq.~(\ref{eq:1LRGI-btorrel-n=1}) we can see that our
procedure has reproduced the correct $1-$loop--renormalization--group
improved behaviour while at the same time fixing the finite parts.
Substituting eq.~(\ref{eq:mHF-btorrel-n=1}) into
eqs.~(\ref{eq:genLL-1-n=1}) and eq.~(\ref{eq:genLL-1-n=1}) we obtain a
manifestly finite form for the parameters
\begin{equation}
  \label{eq:mHF-renpardef-n=1}
  \theta=6\,\Omega^{(2)}=\lambda\,[{\bf
  1}-\tfrac32\,\lambda\,J^{(2)}]^{-1}\quad,\qquad m^2= \tfrac13
  \lambda\,[1 - \tfrac32 \,\lambda\,\tilde J^{(2)}(0)]^{-1}\,v^2
\end{equation}
in terms of $\lambda$ and of the subtracted integral $J^{(2)}$
\begin{equation*}
  J^{(2)}=I^{(2)}-\tilde I^{(2)}(q_s)\,{\bf 1}
\end{equation*}
In particular, recalling the natural definition of the running coupling
constant in eq.~(\ref{eq:run-coupl-n=1}), we see that now
\begin{equation}
  \tilde\theta(p) = \lambda(p) = \frac{\lambda}
  {1 - \tfrac32\,\lambda\,\tilde J^{(2)}(p)} 
\end{equation}
Now, for what concerns RG--invariance, one verifies from
eq.~(\ref{eq:mHF-btorrel-n=1}) that the parameterization of $\lambda_0$
does not depend on the renormalization scale. In fact, renormalizing at
scale $s$ with constant $\lambda=\lambda(q_{s})$ or at scale $s'$ with
constant $\lambda'=\lambda(q_{s'})$ indeed defines the same $\lambda_0$:
\begin{equation*}
  \frac{1}{\lambda_0}=\frac{1}{\lambda}+
  \frac32\,\tilde I^{(2)}(q_s)=\frac{1}{\lambda'}+ 
  \frac32\,\tilde I^{(2)}(q_{s'})
\end{equation*}
Therefore eqs.~(\ref{eq:mHF-pardef-n=1}) can be regarded as manifestly
RG--invariant definitions of the parameters.

Finally let us observe that, since $I^{(2)}$ depends on $m^2$, we have
here an implicit definition of the physical mass by the finite
self--consistent relation in eq.~(\ref{eq:mHF-renpardef-n=1}), rather than
an explicit definition, such as the tree--level
$m^2=\tfrac13\,\lambda\,v^2$, which can be recovered only if $s=0$.

\vskip15pt 
We give now a brief diagrammatic interpretation of the results just
derived. First of all, with the new form of $m^2$ in
eq.~(\ref{eq:mHF-pardef-n=1}), the free propagators are no longer defined
as the resummation of tadpole corrections. In fact are included contributions
from graphs as in fig.. These were actually present in the HF definition of
the external propagator. Moreover contributions from graphs as 
\begin{equation*}
  \centering
  \includegraphics[width=0.40\textwidth]{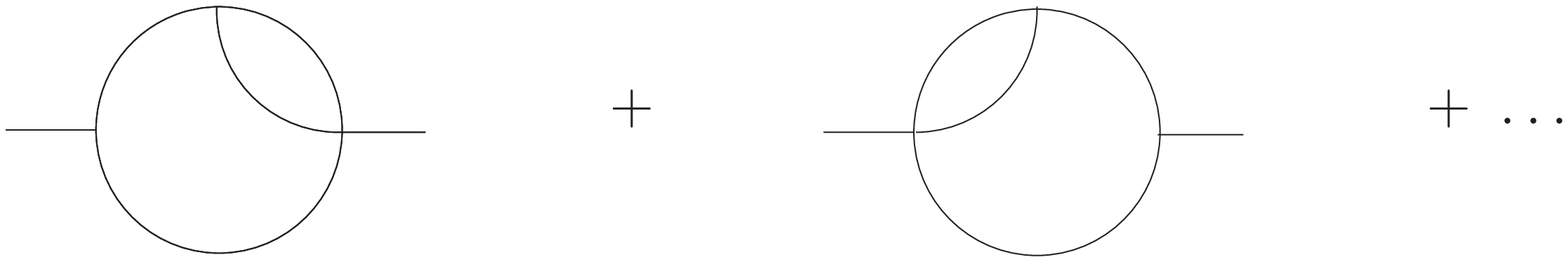}
\end{equation*}
are included in such a way that the $\log \Lambda$ dependence of $m^2$ (at
fixed $\lambda_0$) corresponds to the correct
$1-$loop--renormalization--group improved series.  Notice that corrections
of this type, if included completely, would define a momentum dependent
self--energy. On the other hand it is implicit in a mean--field
approximation that the internal propagators have a momentum--independent
self--energy. Thus the contributions of these diagrams are to be included
in a ``local'' fashion (i.e. with no momentum passing through the loops).

Now, for what concerns the new form of $\Omega^{(2)}$ in
eq.~(\ref{eq:mHF-pardef-n=1}), we can see that it corresponds to the
inclusion, in addition to the chain diagrams of the pure HF approximation,
of all the Leading Log contributions from diagrams as 
\begin{equation*}
  \centering
  \includegraphics[width=0.65\textwidth]{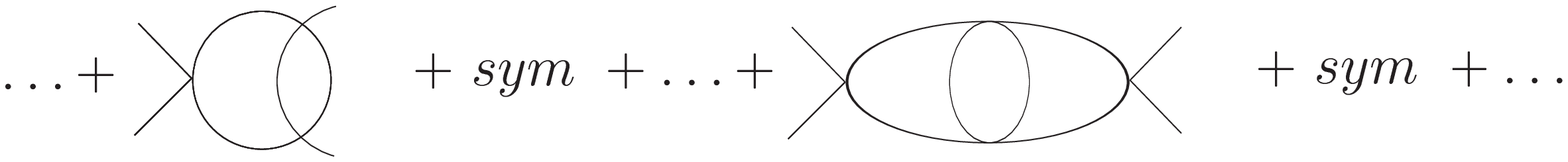}
\end{equation*}
As before their $\log \Lambda$ dependence is taken while their finite
part is fixed by our procedure to be the same of $I^{(2)}$.

In conclusion, the modification of the effective vertex $\theta$ in
eq.~(\ref{eq:mHF-pardef-n=1}) corresponds to include, in the
diagrammatic resummation that defines the $\Omega^{(k)}$ ($k>2$),
Leading Logarithmic contributions from graphs of the form
\begin{equation*}
    \centering
    \includegraphics[scale=0.4]{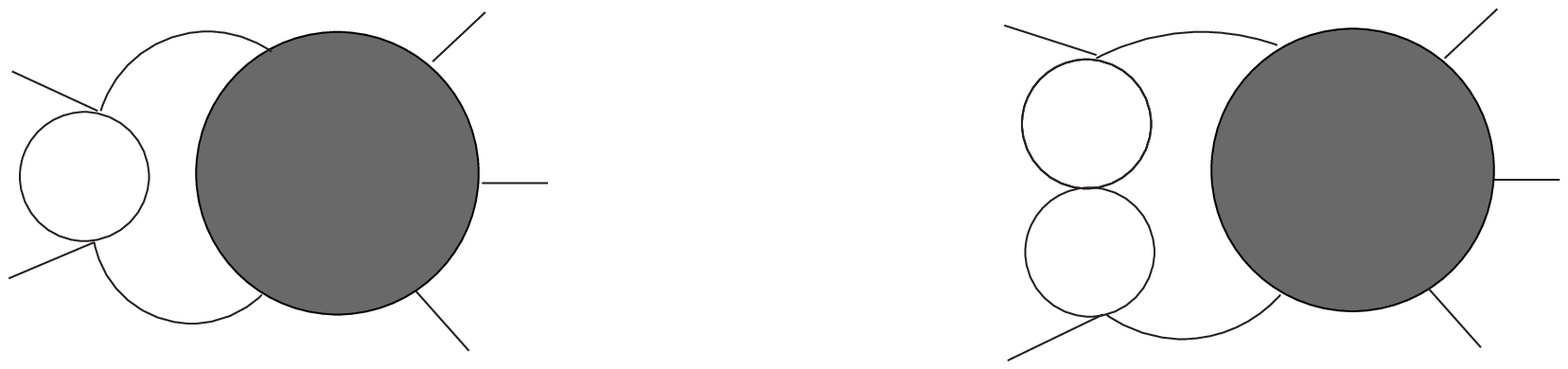}
\end{equation*}
\begin{equation*}
    \centering
    \includegraphics[scale=0.4]{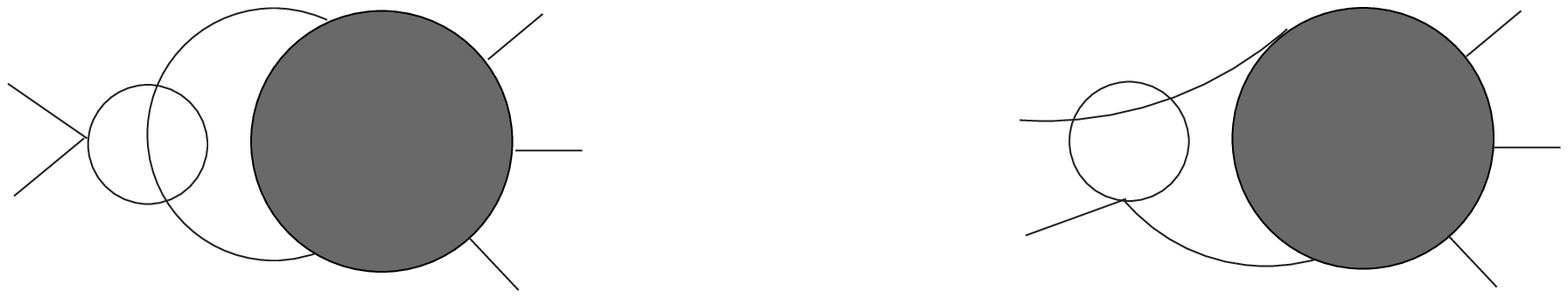}
\end{equation*}
The finite parts of these diagrams are chosen in such a way to
maintain the main Hartree-Fock--like features of the effective vertex,
namely the single channel structure and its form as a function of
$I^{(2)}$.
\vskip20pt
\noindent
Let us make some final observations on the obtained result. The
equations of motion of this modified Hartree-Fock approximation read
\begin{equation}
  \label{eq:mHF-eom-mf}
  \begin{split}
    &\{ \Box + \V(x) - \tfrac13\,\theta\,\Delta\xi(x) \}\phi(x)=0
    \quad,\qquad \{ \Box + \tfrac13\,\tilde\theta(0)\,v^2 + \V(x)
    \} u_{{\bm k}}(x) =0 \\[2mm] &\V = \tfrac12\,\theta\,\{\Delta
    \xi + I - {\hat I}^{(1)}- I^{(2)}\,\V \}
  \end{split}
\end{equation}
As we can see, in order to obtain renormalizability and
RG--invariance, we had to introduce space time non-locality in the
equations of motion. Causality in this nonlocal evolution is
guaranteed by the analyticity in the upper halfplane of $\tilde
I^{(2)}(p)$. Actually in the third equation in
eqs.~(\ref{eq:mHF-eom-mf}) the definition of $\V$ is implicit and it
should be solved for $\V$ in order to have a manifestly causal
form. This can be done paying the price of losing manifest finiteness
\begin{equation*}
  \V = \tfrac12\,\lambda\,[{\bf 1} - \lambda\,( I^{(2)}-
    \tfrac32\,\tilde I^{(2)}(q_s)\,{\bf 1})]^{-1}\,\{\Delta
  \xi + I - {\hat I}^{(1)}\}
\end{equation*}

\subsection{Other initial states} 
\label{sec:other-initial-states}

We already pointed out, in sec.~\ref{sec:analys-hf-appr}, that
eqs.~(\ref{eq:mHF-eom-mf}), with vacuum initial kernels and generic
initial conditions on the background field, already describe an
out--of--equilibrium problem. In this section we study whether and how
we can choose different initial conditions for the mode functions
without spoiling the properties of renormalizability and RG
invariance. We already considered this problem in~\cite{Destri:2005qm}
treating the unbroken symmetry case. The results in this section will be
very similar.

As an example consider an initial state of the same form of the HF
vacuum
\begin{equation*}
  \tilde{\G} ({\bm k} ) = \frac{1}{2 \Omega_{\bm k}} \quad, \qquad
  \tilde{\s} ( {\bm k} )=0 \quad,\qquad \Omega_{\bm k}^2={\bm k}^2 +
  M^2
\end{equation*}
but with a mass $M$ different from the equilibrium mass $m$ defined in
eq.~(\ref{eq:mHF-pardef-n=1}) with $M\neq m$. The corresponding free
mode functions have no longer the plane wave form $u^{(0)}=u^{({\rm vac})}$
[see eq.~(\ref{eq:vac-sol-n=1})]
\begin{equation*} 
  u^{(0)}_{\bm{k}}(x) = \frac{e^{ i \bm{k} \cdot\bm{x}}}{2 \sqrt{2
      \Omega_{\bm k}}} \left\{ \left[ 1 + \frac{\Omega_{\bm
        k}}{\omega_{\bm k}} \right] e^{-i \omega_{\bm k} t}+ \left[ 1 -
    \frac{\Omega_{\bm k}}{\omega_{\bm k}} \right] e^{-i \omega_{\bm k}
    t}\right\}
\end{equation*}
and the free correlation function is not translationally invariant in
time. Then the integral term in eqs.~(\ref{eq:mHF-eom-mf}) 
\begin{equation*} 
  I - \hat I^{(1)} - I^{(2)}\, \V
\end{equation*}
in spite of the subtractions still contains the superficially divergent
contribution
\begin{equation*}
  \int \frac{d^3 k}{(2 \pi)^3}\, \frac{m^2- M^2}{4 \omega_{\bm k}^3}
  \cos{2 \omega_{\bm k} t} 
\end{equation*}
that indeed diverges with the cut--off when $t=0$.  These initial time
singularities can be removed by a Bogoliubov transformation on the
initial state as first observed in \cite{Baacke:1997zz}.
This transformation redefines the initial kernel in
such a way that the leading terms of an high--momentum expansion are
the same as at equilibrium
\begin{equation*}     
  \hat{\G} ( \bm{k} ) \sim \frac{1}{2 \sqrt{{\bm k}^2}}+
  \frac{m^2}{4({\bm k}^2)^{3/2}}+ \dots
\end{equation*}
Initial singularities as well as any other divergence turn to be
absent for any choice of kernel having the above large $\bm k$
expansion. A simple interpretation is that the renormalization
procedure ensures finiteness for any initial Gaussian state belonging
to the same Fock space of the HF vacuum. Extrapolating from this
simple example a generic condition on the short--distance behaviour
of the initial state kernel, we have
\begin{equation*}     
  \hat{\G} ( {\bm x}, {\bm y}) \simeq \frac{1}{4\pi^2|{\bm x}-{\bm y}|}+
  \frac{m^2}{8\pi^2}\log|{\bm x}-{\bm y}| + \dots
\end{equation*}
This ensures the cancellation of all divergent terms are guaranteed by mass and
coupling constant renormalization.

\section{The case $N>1$ }\label{sec:on-case}

We are now ready to consider the more general case of a scalar field
theory with spontaneously broken $O(N)$ symmetry. In doing this we
proceed following closely sec.~\ref{sec:n=1-case}.

Before we begin we fix some notational conventions that will 
help us to handle $O(N)$ index structures while keeping formulas
simple and similar to those in sec.~\ref{sec:n=1-case}. In
particular we will use the standard matrix notation for objects with
one and two indices
\begin{equation}
  \label{eq:matr-molt-ruls}
  \begin{split}
    &[M\,V]_{i}=M_{ij}\,V_{j} \quad,\qquad
    [M^{(1)}\,M^{(2)}]_{ij}=M^{(1)}_{ik}\,M^{(2)}_{kj} \\&{\bf
    1}_{ij}=\delta_{ij} \quad,\qquad [V\,V^T]_{ij}=V_i \,V_j
    \quad,\qquad V^2=V_i\,V_i
  \end{split}
\end{equation}
When objects with four indices are concerned, we will use the
following conventions 
\begin{equation}
  \label{eq:tens-molt-ruls}
  \begin{split}
    &[T\,M]_{ij}=T_{ijkm}M_{km} \quad,\qquad
    [T^{(1)}\,T^{(2)}]_{ijkm}=T^{(1)}_{ijrs}\,T^{(2)}_{rskm} \\[2mm]
    &\Pi_{ijkm}=\tfrac12
    (\delta_{ik}\delta_{jm}+\delta_{im}\delta_{jk})\quad,\qquad
    T^{-1}\,T=T\,T^{-1}=\Pi
  \end{split}
\end{equation}
where the last relation is regarded as restricted to tensors with the
symmetry $T_{ijkm}=T_{ijmk}=T_{jikm}$. Moreover we will use also the
following definitions
\begin{equation*}
  {\hat {\bf 1}}(x,y) = {\bf 1}\, \delta^{(4)}(x - y)
  \quad,\qquad {\hat \Pi}(x,y)=\Pi\,\delta^{(4)}(x - y)
\end{equation*}

\subsection{Analysis of the HF approximation}
\label{sec:analys-hf-appr-1}

The Hartree-Fock equations of motion for a scalar $O(N)$ theory are
shown in eqs.~(\ref{eq:HF-eom-mf}). As a first example of the use of
the notational conventions in eq.~(\ref{eq:matr-molt-ruls}) and
eq.~(\ref{eq:tens-molt-ruls}) we rewrite them in a compact indiceless
form
\begin{equation}
\label{eq:hf-eom-on}
  \begin{split} 
    &\left\{[\,\Box + m_0^2 +\tfrac16\lambda_0 \,{\rm tr}\,\xi\,]
      \,{\bf 1} + \tfrac12 \lambda_0\,\tau\, 
      \int_{p^2<\Lambda}\frac{d^3 p}
      {(2\pi)^3}~ u_{\bm{p}}\,u_{\bm{p}}(x)^\dag \right\} \phi=0
    \\[2mm] & \left\{[\,\Box + m_0^2\,]\,{\bf 1} +
      \tfrac12\,\lambda_0\,\tau \,\left[\xi +\int_{p^2<\Lambda}
        \frac{d^3 p }{(2\pi)^3}~u_{\bm{p}}\,u_{\bm{p}}^\dag 
      \right]\,\right\} u_{\bm{k}}=0
  \end{split}
\end{equation}
where we recall that $\xi (x)= \phi(x)\,\phi^T(x)$ and the definition
of $\tau$ in eq.~(\ref{eq:tau-tens-def}) .

Exactly as in sec.~\ref{sec:analys-hf-appr} we select the values of bare
parameters corresponding to the broken symmetry phase by requiring the
existence of a vacuum solution with non zero constant and uniform
background field, $\phi(x)=v$. Notice that here $v$ is a $O(N)$. The
direction ${\hat v}=v/\sqrt{v^2}$ of the vacuum background field provides
the definition of longitudinal and transverse projectors
\begin{equation*}
  P_L={\hat v}\,{\hat v}^T \qquad, \quad P_T={\bf 1}-P_L
\end{equation*}
The vacuum mode functions now are
\begin{equation*}
  \begin{split}
    u^{({\rm vac})}_{{\bm k}}(x) &= (2\, \omega_{\bm k})^{-1/2}
    \,\exp[i(\bm{k}\cdot{\bm x})\,{\bf 1}-i\omega_{\bm k} t]
    \\[2mm] &= \left[ (2\,\omega_{\bm{k},L})^{-1/2}\,
    e^{-i\omega_{\bm k,L}\,t}\,P_L + (2\, \omega_{\bm{k},T})^{-1/2}\, 
    e^{-i\omega_{\bm k,T}\,t}\,P_T \right]\,e^{i\,\bm{k}\cdot{\bm x}}
  \end{split}
\end{equation*}
where in the first line the power and the exponentiation are operations on
matrices and $\omega$, $\omega_L$ and $\omega_T$ are defined as
\begin{equation*}
  \omega^2_{\bm k} =\omega^2_{\bm k,L}\,P_L +\omega^2_{\bm k,T}\,P_T =
  {\bm k}^2\,{\bf 1} + M^2 \quad , \qquad M^2=m_L^2\,P_L + m_T^2\,P_T
\end{equation*}
The longitudinal mass has the same value as in the $N=1$ case, namely
$m_L^2=\tfrac13\lambda_0 v^2$, while the transverse squared mass $m_T^2$,
as a function of $\lambda_0$, $v^2$ and $\Lambda$, is obtained by solving
the self--consistent equation
\begin{equation}
  \label{eq:m_T-def-long}
  0=m^2_T + \tfrac13\,\lambda_0\,\int_{p^2<\Lambda} \frac{d^3 p}{(2
    \pi)^3} \,\left\{ \frac{1}{2\, \omega_{\bm k,L}}- \frac{1}{2\,
    \omega_{\bm k,T}}\right\}
\end{equation}
The vacuum expectation value $v$ satisfies a gap equation which generalizes
the one in eq.~(\ref{eq:gap-eq-n=1})
\begin{equation}
\label{eq:gap-eq-on}
    0 = m^2_0 + \tfrac16\,\lambda_0\,v^2 + \tfrac12 \,\lambda_0
    \int_{p^2<\Lambda} \frac{d^3 p}{(2 \pi)^3} \,\frac{1}{2\,
    \omega_{\bm k,L}}+ \tfrac16(N-1) \,\lambda_0 \int_{p^2<\Lambda}
    \frac{d^3 p}{(2 \pi)^3} \,\frac{1}{2\, \omega_{\bm k,T}}
\end{equation} 
Notice that the gap equation involves only $v^2$. In fact varying the
direction of $\hat v$ we obtain different static solutions that
correspond to the equivalent distinct vacua of the theory in the
broken symmetry phase.
\vskip 15pt
\noindent
The mean field $\V$ [see eq.~(\ref{eq:V-def-n=1})] is now an object
with two $O(N)$ indices, defined by
\begin{equation*}
  \V(x) = m_0^2\,{\bf 1} - M^2 + \tfrac12 \lambda_0 \,\tau \big[
    \xi (x) + \int_{p^2<\Lambda} \frac{d^3 p }{(2\pi)^3}~
    |u_{\bm{p}}(x)|^2 \big]
\end{equation*}
Notice that again $\V=0$ on the vacuum solution. In terms of $\V$
eqs.~(\ref{eq:hf-eom-on}) read
\begin{equation*}
  \begin{split} 
    &\left\{\Box\,{\bf 1} + M^2 - \tfrac13 \lambda_0\,\xi + \V
      \right\} \phi(x)=0 \\[2mm] & \left\{\Box \,{\bf 1} + M^2
      +\V \,\right\} u_{\bm{k}}(x) =0
  \end{split}
\end{equation*}
Recalling the general mean field form of the background field
equation ($\F'$ here is a $O(N)$ matrix)
\begin{equation}
  \label{eq:gen-bkeq-on}
  \frac{\delta \Gamma_{\rm 1PI}}{\delta \phi_{\Delta}}
  \Bigg|_{\substack{\phi_\Delta = 0 \\\phi_c = \phi}} =(\Box
  \, {\bf 1}+ 2 \,\F'[\xi])\, \phi=0
\end{equation}
we have 
\begin{equation*}
   \F'[\xi]= \tfrac12\,\V+\tfrac12\,M^2 -\tfrac16\,\lambda_0\,\xi
\end{equation*}
which should be compared with eq.~(\ref{eq:FV-rel-n=1}) in
sec.~\ref{sec:analys-hf-appr}.  

\vskip15pt
In order to derive a self--consistent equation for $\V$ like the one in
eq.~(\ref{eq:scon-eq-n=1}), as in sec.~\ref{sec:analys-hf-appr} we
introduce the free mode functions
\begin{equation*}
  ( \Box + M^2 )u_{{\bm k}}^{(0)} ( x ) = 0\quad,\qquad u_{{\bm
      k}}^{(0)}(\bm{x}, 0) = u_{{\bm k}}(\bm{x}, 0)\quad,\qquad \dot
  u_{{\bm k}}^{(0)}(\bm{x}, 0) = \dot u_{{\bm k}}(\bm{x}, 0)
\end{equation*}
and define the free propagators
\begin{equation}
  \label{eq:free-prop-on}
  \begin{split}
    &G_R^{(0)} (x-y) = G_{R,L}^{(0)} (x-y)P_L+G_{R,T}^{(0)} (x-y)P_T= 
    \int \frac{d^4 p }{(2 \pi)^4}\, \frac{e^{-i p\cdot(x-y)}}{(p^2+ i 
      \epsilon p_0)\,{\bf 1} - M^2} \\[2mm] 
    &G_{A,ij}^{(0)} (x-y) = G_{R,ji}^{(0)} (y-x) \\[2mm]  &G_{ij}^{(0)} 
    (x,y) = G_L^{(0)} ( x ,y )P_{L,ij}+G_T^{(0)} ( x,y )P_{T,ij}=
    \mathrm{Re} \int_{|\bm p|<\Lambda} \frac{d^3 p }{(2
    \pi)^3}\,u_{\bm{p},i a}^{(0)}(x)\, \overline{u}_{\bm{p},a
    j}^{(0)}(y)
  \end{split}
\end{equation}
Performing the same manipulations on the mode function equations as in
sec.~\ref{sec:analys-hf-appr} we can rewrite the cutoffed
correlation, that now reads
\begin{equation*}
  G_{ij} (x,y) = \mathrm{Re} \int_{|\bm p|<\Lambda} \frac{d^3 p }{(2
    \pi)^3} \, u_{\bm{p},i a}(x)\, \overline{u}_{\bm{p}, a j}(y)
\end{equation*}
as a functional of $\V$
\begin{equation*}
  G = G[{\cal V}] = [{\hat {\bf 1}}-G_R^{(0)}\hat{\cal V}]^{-1}
  \,G^{(0)}\,[{\hat {\bf 1}}-\hat{\cal V} G_A^{(0)}]^{-1}
\end{equation*}
In conclusion we have the self--consistent equation for $\V$ 
\begin{equation}
  \label{eq:scon-eq-on}
  \V=m_0^2\,{\bf 1}- M^2 + \tfrac12\,\lambda_0\,\tau\,(\xi + I[\V])
  \quad,\qquad I[\V](x) \equiv G[\V](x,x)
\end{equation}
This should be compared with eq.~(\ref{eq:I-def-n=1}) and
 eq.~(\ref{eq:scon-eq-n=1})

\vskip15pt
We keep on following closely the sec.~\ref{sec:analys-hf-appr} by
fixing vacuum initial conditions for the mode
functions. By this choice follows that $u^{(0)}=u^{({\rm vac})}$ for all
times and the free correlation is translationally invariant.  

Moreover we can repeat, with little changes, the observations made in
sec.~\ref{sec:analys-hf-appr} about the structure of the integral
$I[\V]$. We can still interpret the free Green functions as effective
internal propagators obtained by resumming all tadpole corrections to
the bare ones. We can introduce the tadpole integral
\begin{equation*}
  I^{(1)}= G^{(0)}(x,x)=  I_L^{(1)}\,P_L+I_T^{(1)}\,P_T
\end{equation*}
and the two propagators loop integral
\begin{equation}
  \label{eq:2prop-int-on}
  \begin{split}
    I_{ijkm}^{(2)}(x-y) &=G^{(0)}_{R,ik}(x-y)\, G^{(0)}_{mj}(x,y)+
    G^{(0)}_{R,mj}(x-y)\, G^{(0)}_{ik}(x,y)=\\
    &=I_{LL}^{(2)}(x-y)\,P_{L,ik}\,P_{L,jm}+
    I_{TT}^{(2)}(x-y)\,P_{T,ik}\,P_{T,jm}+\\
    &+\tfrac12\,I^{(2)}_{TL}(x-y)\,(\,P_{L,ik}\,P_{T,jm}+P_{T,ik}\,P_{L,jm})
  \end{split}
\end{equation}
In terms of these we can define a functional $J$ by subtracting to $I$
its constant part and the term linear in $\V$. By the same arguments
used in sec.~\ref{sec:analys-hf-appr} we can see that $J$ contains
only the cut-off convergent part of $I$. The definition of the $J$
functional is formally unchanged respect to eq.~(\ref{eq:J-def-n=1})
[here $O(N)$ indices contraction according to the rules in
eq.~(\ref{eq:tens-molt-ruls}) is understood]
\begin{equation}
  \label{eq:J-def-on}
  J[\V] = I[\V]- I^{(1)} - I^{(2)}\,\V
\end{equation}
Notice also that, by using the two propagator loop integral of
eq.~(\ref{eq:2prop-int-on}), we can cast the self--consistent
definition of $m^2_T$ in eq.~(\ref{eq:m_T-def-long}) into a more
compact form
\begin{equation}
  \label{eq:eq:m_T-def-comp}
  m^2_T= - \tfrac13\,\lambda_0\,\tilde I_{TL}^{(2)}(0)\,m^2_T\,[
    1-\tfrac13\,\lambda_0 \,\tilde
    I^{(2)}_{TL}(0)]^{-1}
\end{equation}
The ``quasi-renormalized'' form is obtained by manipulations identical
to those in sec.~\ref{sec:analys-hf-appr}
\begin{equation*}
  \V= \tfrac12 \,\theta\, \Delta \xi + \Delta \V \; , \quad \Delta
  \V =\tfrac12 \,\theta \, J[\Delta \V + \tfrac12\,\theta\,\Delta
  \xi]\;,\quad \F'=\tfrac12\,\Omega^{(1)}+\tfrac12\,\Omega^{(2)}\,\Delta \xi
  + \tfrac12\,\Delta \V
\end{equation*}
The matrix $\Omega^{(1)}$ and the four--indices objects $\Omega^{(2)}$
and $\theta$ are defined as
\begin{equation}
  \label{eq:thet-Ome-def}
  \Omega^{(1)}= m_T^2\,P_T \quad ,\qquad \theta = \lambda_0 \, 
  \left[ {\hat\Pi} - \tfrac12\,\lambda_0 \, I^{(2)} \right]^{-1}
  \quad, \qquad \Omega^{(2)} = \tfrac12 \, \theta - \tfrac13\,
  \lambda_0\,{\hat\Pi}
\end{equation}

The vertex functions contributing to the equation of motion [see
eq.~(\ref{eq:gamma-CPT-n=1})] are obtained from the variation of
eq.~(\ref{eq:gen-bkeq-on}) with respect to $\phi_c$ at
$\phi=v$. They can be expressed in terms of functional derivatives of $\F'$
\begin{equation*}
  \Omega^{(k)}_{i_1 j_1 \dots i_k j_k}(x_1,\dots,x_{k})=
  2\,\frac{\delta\F'[\xi](x_1)}{\delta \xi_{i_1,j_1}(x_2)\dots \delta
    \xi_{i_k j_k}(x_k)}\Big|_{\xi=v^2}
\end{equation*}
as in eq.~(\ref{eq:Gamma-Omega-n=1}).  Notice that $\Omega^{(1)}$ is
completely transverse [see eq.~(\ref{eq:thet-Ome-def})] which implies that
$\Gamma^{(1)}=\Omega^{(1)}v=0$, namely the statement that the vector $v$ is
the vacuum static solution.

Of course all the diagrammatic interpretations made in
sec.~\ref{sec:analys-hf-appr} still hold true, now with diagrams carrying
$O(N)$ indices.  

\vskip15pt
Before going further, we should introduce some notations that will be
useful later on. In fact we will have to deal with four--indices
tensors which have some particular symmetry properties and are
functions of $v$ (and of other scalar quantities)
\begin{equation*}
  T_{ijkm}(v)= T_{jikm}(v)= T_{ijmk}(v)= T_{kmij}(v) 
\end{equation*}
as, for example, $\theta$ and $\Omega^{(2)}$. Such tensorial objects admit
a general decomposition
\begin{equation}
  \label{eq:dec-rul-on}
  T_{ijkm}(v)= T_\alpha (v^2) \; t^{\alpha}_{ijkm}({\hat v})\qquad
  \alpha=1,\dots,5
\end{equation}
in terms of the five elementary tensors 
\begin{equation*}
  \begin{split}
    t^1_{ijkm} &= \hat{v}_i\hat{v}_j\hat{v}_k\hat{v}_m 
    \;,\quad ~~t^2_{ijkm}= \tfrac12
    \,(P_{T,ik}\,P_{T,jm}+P_{T,im}\,P_{T,jk})\\ t^3_{ijkm}
    &=P_{T,ij}\,P_{T,km} \;,\quad t^5_{ijkm}=
    (P_{L,ij}\,P_{T,km}+P_{T,ij}\,P_{L,km})\\ t^4_{ijkm}&=
    \tfrac12 \, (P_{L,ik}\,P_{T,jm}+P_{L,im}\,
    P_{T,jk}+P_{T,ik}\,P_{L,jm}+P_{T,im}\,P_{L,jk})
  \end{split}
\end{equation*}
Notice that the coefficients of the decomposition are functions of $v^2$
alone.  As an example, by eq.~(\ref{eq:thet-Ome-def})the coefficients
$\theta_\alpha$ of the decomposition of $\theta$ are
\begin{equation}
  \begin{split}
    \label{eq:HF-theta-comps}
    \theta_1&= \lambda_0\,[{\bf 1}-
      \tfrac19(N+2)\,\lambda_0\,I^{(2)}_{TT}]\,\{ {\bf 1} -
      \tfrac16(N+1)\,\lambda_0\,I^{(2)}_{TT} -
      \tfrac12\lambda_0\,I^{(2)}_{LL}\,[{\bf 1}-
      \tfrac19(N+2)\,\lambda_0\,I^{(2)}_{TT}]\,\}^{-1}\\[2mm] 
      \theta_2 &= \tfrac23 \lambda_0\,[{\bf 1}-
      \tfrac13\,\lambda_0\,I_{TT}^{(2)}]^{-1} \;,\quad 
      ~~~~~~~~~~\theta_4 = \tfrac23 \lambda_0\,[{\bf 1}-
      \tfrac13\,\lambda_0\,I_{TL}^{(2)}]^{-1}\\[2mm] 
      \theta_5&= \tfrac13 \, \theta_1 [{\bf 1} - \tfrac19(N+2)\,
      \lambda_0\,I_{TT}^{(2)}]^{-1}\;,\quad 
      \theta_3=\theta_5\,[{\bf 1} - \tfrac13\,\lambda_0 \,I_{LL}^{(2)}]
      \,[{\hat 1} - \tfrac13\,\lambda_0\,I_{TT}^{(2)}]^{-1}
  \end{split}
\end{equation}
Notice that using eqs.~(\ref{eq:HF-theta-comps}) the compact form of
the self--consistent definition of $m_T^2$ in
eq.~(\ref{eq:eq:m_T-def-comp}) can be rewritten as
\begin{equation}
  \label{eq:eq:m_T-def-theta}
  m^2_T = \tfrac13\,\lambda_0\,v^2 - \tfrac12\,\tilde\theta_4(0)\,v^2
\end{equation} 
The coefficient of $\Omega$ can be determined from those of $\theta$ by
eq.~(\ref{eq:thet-Ome-def}) and the decomposition of the $\Pi$ tensor
\begin{equation}
  \label{eq:HF-Om2-comps}
  \begin{split}
    \Omega_\alpha^{(2)}&=\tfrac12\,\theta_\alpha -
    \tfrac13\,\lambda_0\,{\bf 1}\quad,\qquad \alpha=1,2,4\\
    \Omega_\alpha^{(2)}&=\tfrac12\,\theta_\alpha \quad,\qquad
    \alpha=3,5\\
  \end{split}
\end{equation}

\subsection{Renormalization}
\label{sec:renorm-rg-invar-1}
Now we apply the standard renormalization procedure as we did in
sec.~\ref{sec:renorm-rg-invar} for the $N=1$ case. We will see that the
same problems are present also in this case and some others will arise due
the presence of the spontaneously broken continuous $O(N)$ symmetry.

By fixing the equilibrium value of the background field we provide the
first renormalization condition that defines the bare mass $m_0^2$ as a
function of $\lambda_0$, $v$ and $\Lambda$
\begin{equation*}
  m^2_0(\lambda_0,v,\Lambda)= - \tfrac16\,\lambda_0\,v^2 - 
  \tfrac12 \,\lambda_0
  \int_{p^2<\Lambda} \frac{d^3 p}{(2 \pi)^3} \,\frac{1}{2\,
    \omega_{\bm k,L}}+ \tfrac16(N-1) \,\lambda_0 \int_{p^2<\Lambda}
  \frac{d^3 p}{(2 \pi)^3} \,\frac{1}{2\, \omega_{\bm k,T}}
\end{equation*}
We recall that $m_L^2=\tfrac13\,\lambda_0\,v^2$ and $m_T^2$ is given
(in function of $\lambda_0$, $v$ and $\Lambda$) as solution
eq.~(\ref{eq:m_T-def-long}) . Again, enforcing this condition is
enough to remove all $\Lambda^2$ dependence from the vertex functions.

The second renormalization condition is conventionally obtained by
evaluating at the symmetric point [see eq.~(\ref{eq:simm-point})] the
``all--longitudinal'' component of the four--legs vertex function
$\Gamma^{(4)}$ and requiring it to be equal to the renormalized
coupling $\lambda$
\begin{equation*}
  \lambda=\tilde
    \Gamma_{ijkm}^{(4)}(-\tfrac32\,q_s,\tfrac12\,q_s,\tfrac12\,q_s)\,{\hat
    v}_i\,{\hat v}_j\,{\hat v}_k\,{\hat v}_m
\end{equation*}
Explicitly expressed in terms of the variations of $\F'$ the above
expression involves the (longitudinal components) of $\Omega^{(2)}$,
$\Omega^{(3)}$ and $\Omega^{(4)}$. As in
sec.~\ref{sec:renorm-rg-invar} [see eq.~(\ref{eq:recon-coumod-n=1})]
we drop the contributions from $\Omega^{(3)}$ and $\Omega^{(4)}$
obtaining a simpler definition
\begin{equation}
\label{eq:recon-coumod-on}
  \lambda=6\,\tilde \Omega_{ijkm}^{(2)}(q_s)\,{\hat v}_i\,{\hat
    v}_j\,{\hat v}_k\,{\hat v}_m =6\,\tilde \Omega_1^{(2)}(q_s)
\end{equation}
or, thanks to eq.~(\ref{eq:thet-Ome-def}),
\begin{equation}
\label{eq:HF-btorrel-on}
  \lambda=\frac{3\,\lambda_0\,[1 -
      \tfrac{N+2}9\,\lambda_0\,\tilde I^{(2)}_{TT}(q_s)]}{1 -
      \tfrac{N+1}6\,\lambda_0\,\tilde I^{(2)}_{TT}(q_s) -
      \tfrac12\lambda_0 \tilde I_{LL}^{(2)}(q_s)\,[1 -
      \tfrac{N+2}9\,\lambda_0\,\tilde
      I^{(2)}_{TT}(q_s)]}-\tfrac13\,\lambda_0
\end{equation}
This has to be compared with the corresponding relation of the $N=1$
case in eq.~(\ref{eq:HF-btorrel-n=1}).

\vskip15pt
First of all eq.~(\ref{eq:HF-btorrel-on}) shows the same pathological
dependence of $\lambda$ on $\lambda_0$, at fixed $\Lambda$
and $v$, that we found in sec.~\ref{sec:renorm-rg-invar}. As already
discussed this prevents a consistent map between bare parameters and
renormalized ones unless we restrict to a small coupling regime. The
correct 1--Loop--Renormalization--Group improved relation, instead, reads
\begin{equation}
  \label{eq:1LRGI-btorrel-on}
  \frac{1}{\lambda} = \frac{1}{\lambda_0} + \frac{N+8}{48 \pi^2} \,\log
  \Lambda + \dots
\end{equation}
where again the dots stand for some suitable choice of the renormalization
scale and of the finite parts.

For what concerns the cut--off dependence of $\Omega^{(2)}$ we can see
that things are even worse than in the $N=1$ case. In fact, not only
the $\log \Lambda$ dependence is not removed by the renormalization
condition from $\tilde \Omega^{(2)}(p)$ for $p\neq q_s$, but also, for
components $\Omega_\alpha$ with $\alpha \neq 1$ logarithmic
divergences already appear at $p=q_s$ as can be verified by the
explicit definitions in eqs.~(\ref{eq:HF-Om2-comps}).

Regarding the cut--off dependence of the $\Omega^{(k)}$ with $k>2$ the same
observations of sec.~\ref{sec:renorm-rg-invar} still hold in this case. In
particular we still have $\log \Lambda$ dependence in the masses of the
free propagators. This is immediate for $m_L^2=\tfrac13\,\lambda_0\,v^2$
and can be easily verified for $m_T^2$ by considering its self--consistent
definition in eq.~(\ref{eq:m_T-def-long}). Moreover, as in the $N=1$ case,
$\log \Lambda$ divergences persist in the effective vertex $\theta$ after
imposing the renormalization condition [eq.~(\ref{eq:recon-coumod-on})].

\vskip15pt
Besides these renormalization problem analogous to the $N=1$ case, another
very important aspect, which is peculiar of the $N>1$ case, must be
pointed out. First of all, we may verify the formal $O(N)$ invariance of
the HF approximation also in the current out--of--equilibrium context. In
fact, eqs.~(\ref{eq:hf-eom-on}) are manifestly $O(N)$ symmetric and so it
should be for the effective action
\begin{equation*}
  \Gamma_{1PI}[\phi_\Delta,\phi_c]=\Gamma_{1PI}[R\,\phi_\Delta,R\,\phi_c]
  \quad, \qquad \forall R \in O(N)
\end{equation*}
Hence for infinitesimal $R={\bf 1} + \epsilon = {\bf 1} - \epsilon^T$,
to first order we must have
\begin{equation*}
  0= \int~d^4x \left\{\phi_\Delta^T(x)\,\epsilon\,
  \left(\frac{\delta}{\delta
  \phi_\Delta(x)}\,\Gamma_{1PI}[\phi_\Delta,\phi_c]\right)+\phi_c^T(x)\,
  \epsilon \, \left(\frac{\delta}{\delta
  \phi_c(x)}\,\Gamma_{1PI}[\phi_\Delta,\phi_c]\right)\right\}
\end{equation*}
Upon variation with respect to $\phi_\Delta$ at $\phi_\Delta=0$, $\phi_c=v$
and recalling that $\Gamma^{(1)}=0$, we conclude
\begin{equation*}
  \int~d^4x \, v_i\,\epsilon_{ij}\,\Gamma_{jk}^{(2)}(x-y)=0\quad 
  \Longrightarrow \quad \tilde \Gamma_T^{(2)}(0)=0
\end{equation*}
This Ward identity, stating the masslessness of the external
transverse propagator (the inverse of $\tilde{\Gamma}^{(2)}_{T}(p)$), is
indeed satisfied by the HF approximation
\begin{equation}
  \label{eq:HF-2legs-on}
  \begin{split} 
    & \Gamma^{(2)}=\Gamma^{(2)}_{L}\,P_L + \Gamma^{(2)}_{T}\,P_T
    \\[2mm] &\tilde{\Gamma}^{(2)}_{L}(p)= - p^2 + 2 \,
    \tilde\Omega_1^{(2)}(p)\, v^2\\[2mm]
    &\tilde{\Gamma}^{(2)}_{T}(p)= - p^2 + \Omega^{(1)}_T + 
    \tilde\Omega^{(2)}_4 (p)\, v^2= - p^2 + m^2_T + \left[\tfrac12 \,
      \tilde\theta_4
      (p)- \tfrac13 \, \lambda_0 \right] v^2
  \end{split}
\end{equation}
thanks to the self--consistent definition of the transverse mass,
eq.~(\ref{eq:eq:m_T-def-theta}). But the transverse mass itself, which
enters the internal transverse propagators $G^{(0)}_{R,T}$ and $G^{(0)}_T$
does not vanish at all (it actually diverges as $\log\Lambda$), preventing
a consistent interpretation of the transverse modes as Goldstone bosons.
This is a well known problem of the HF approximation (see for instance
ref.~\cite{vanHees:2002bv}).

\subsection{A class of improved HF approximations} 
\label{sec:defin-modif-hf-1}

We shall now try and improve the HF approximation to recover the correct
properties of Renormalizability and RG-invariance. We shall also require
that this improved resummation is gapless; that is, we shall impose that
the internal transverse propagators are massless.  To this end we follow
closely the procedure of sec.~\ref{sec:defin-modif-hf}, while stressing
some important new features which appear due to the continuous $O(N)$
symmetry and the presence of two kinds of fields (transverse and
longitudinal) with different masses. Let us anticipate that in this case
our procedure does not single out a unique solution but rather an extended
class of resummations that share all the required properties.

As in sec.~\ref{sec:defin-modif-hf} the first step consists in fixing some
fundamental properties of the HF approximation that we want to preserve.
First of all the general structure of the diagrams resummation which is
encoded in the general mean field form of the background equations of
motion:
\begin{equation}
  \label{eq:gen-feat1-on}
  \frac{\delta \Gamma_{\rm 1PI}}{\delta \phi_{\Delta,i}
    (x)}\Bigg|_{\substack{\phi_\Delta = 0 \\\phi_c = \phi}} =\big( \Box
  \, {\bf 1}+ 2 \,\F'[\xi](x)\big)_{ij} \phi_j(x)=0  
\end{equation}
and in the self--consistent definition of $\F'$
\begin{equation}
  \label{eq:gen-feat2-on}
  \F'=\tfrac12\,\Omega^{(1)}+[\tfrac12\,\Omega^{(2)}-\tfrac14\,\theta]\,\Delta
  \xi + \tfrac12\, \V\quad,\qquad \V= \tfrac12 \,\theta\, \Delta \xi +
  \tfrac12 \,\theta \, J[\V]
\end{equation}
which imply the following mean--field--type equations of motion for the
mode functions
\begin{equation}
  \label{eq:gen-mfeq-on}
  \left[ \Box {\bf 1} + M^2 + \V \right] u_{{\bm k}}
  =0 \quad,\qquad \V = \tfrac12\,\theta\,\{\Delta \xi + I - 
    I^{(1)}- I^{(2)}\,\V \}
\end{equation}
As in sec.~\ref{sec:defin-modif-hf} we now regard $\Omega^{(1)}$,
$\Omega^{(1)}$, $\theta$ and the free propagators masses $m_L^2$ and
$m_T^2$ (in matrix form $M^2$) as tunable parameters that we are going to 
change w.r.t. to their HF definitions
\begin{equation}
  \label{eq:HF-pardef-on}
  \begin{split}
    &\theta = \lambda_0 \, \left[ {\hat \pi} - \tfrac12\,\lambda_0 \,
      I^{(2)} \right]^{-1} \quad, \qquad \Omega^{(1)}= m_T^2\,P_T
    \quad ,\qquad \Omega^{(2)} = \tfrac12 \, \theta - \tfrac13\,
    \lambda_0\,{\hat \pi}\\[2mm] &m_L^2=\tfrac13\,\lambda_0\,v^2\quad,
    \qquad m^2_T=\tfrac13\,\lambda_0\,v^2-
    \tfrac12\,\tilde\theta_4(0)\,v^2
  \end{split}
\end{equation}
The first important difference with the $N=1$ case is that now
$\Omega^{(2)}$, $\theta$ and $M$ are not independent, but must fulfill
certain $O(N)$ symmetry constraints. In fact the form of the equations of
motion in eqs.~(\ref{eq:gen-feat1-on}) and~(\ref{eq:gen-mfeq-on}) is not
manifestly $O(N)$ symmetric for generic tunable parameters. More precisely
they are covariant under contemporaneous rotations of the field $\phi$, of
the mode functions $u_{\bm p}$ {\em and} of the vacuum expectation value
$v$. The latter, in fact, enters in the definitions the propagators (see
the decomposition in eq.~(\ref{eq:free-prop-on})) and of $\Omega^{(2)}$ and
$\theta$ (see the general decomposition rules in
eq.~(\ref{eq:dec-rul-on})). We have to explicitly require symmetry under
rotations of $\phi$ and $u_{\bm p}$ alone or, equivalently, invariance
under rotations of $v$. In other words we must impose the proper $O(N)$
Ward identities.

Let us therefore define the operator generating the infinitesimal variation
under an $O(N)$ rotation of $v$
\begin{equation*}
  \delta = \epsilon_{ij}\,v_i\,\frac\delta{\delta v_j}
\end{equation*}
where $\epsilon$ is an arbitrary antisymmetric matrix. It is immediate to
verify that it annihilate any invariant function.  

First of all, to require invariance in the mode functions equations
(\ref{eq:gen-mfeq-on}), we have to impose
\begin{equation}
  \label{eq:sym-cond-1.1}
  \delta[M^2 + \V]=0
\end{equation}
Solving explicitly for $\V$ the second equation in
eqs.~(\ref{eq:gen-mfeq-on}) we can rewrite the above condition as
\begin{equation}
  \label{eq:sym-cond-1.2}
  0=\delta M^2 + \delta\{\,[{\bf 1}+
    \tfrac12\,\theta\,I^{(2)}]^{-1}\,\theta\,[\Delta \xi + I - 
    I^{(1)}]\}
\end{equation}
Notice that $\xi$ and $u_{\bm k}$ are independent and by hypothesis
invariant variables (i.e. they must depend only on $v^2$). Then
eq.~(\ref{eq:sym-cond-1.2}) implies
\begin{equation*}
  \label{eq:sym-cond-1.3}
  \delta \{[{\bf 1}+ \tfrac12\,\theta\,I^{(2)}]^{-1}\,\theta\}=0
  \quad,\qquad 0=\delta M^2 + [{\bf 1}+
    \tfrac12\,\theta\,I^{(2)}]^{-1}\,\theta\,[- v^2\, \delta P_L - 
    \delta I^{(1)}]\} 
\end{equation*}
By some easy algebraic manipulations we can rewrite the first equation
above as
\begin{equation*}
  \delta \theta = \tfrac12\,\theta\,\delta I^{(2)}\,\theta
\end{equation*}
which has a simple and natural diagrammatic interpretation. Making the
tensorial structure explicit, we obtain the following conditions on the
coefficients $\theta_\alpha$
\begin{equation}
  \label{eq:par-rel-1}
  \begin{split}
    \theta_2&= \theta_4 \,[{\bf 1} - \tfrac12\,( I_{TT}^{(2)}- 
      I_{TL}^{(2)} \,)\,]^{-1}\\[2mm] \theta_5 &= [\theta_1 - \theta_4
      + \tfrac12 \, \theta_4 \, \theta_1 \,(\, I_{LL}^{(2)}-
      I_{TL}^{(2)} )\,]\,[{\bf 1} + \tfrac12 \, \theta_4
      \,(\,I_{TT}^{(2)} - I_{TL}^{(2)} \,)\,]^{-1}\\[2mm] \theta_3 &=
    \theta_5 \, [{\bf 1} + \tfrac12 \, \theta_4 \,(\, I_{LL}^{(2)}-
      I_{TL}^{(2)}\,)\,][{\bf 1} + \tfrac12 \, \theta_4 \,(\,
      I_{TT}^{(2)}- I_{TL}^{(2)}\,)\,]^{-1}
  \end{split}
\end{equation}
Notice that these relations uniquely fix the form of $\theta$ once
$\theta_1$ and $\theta_4$ have been specified.  Now, using these relations
and the following rules for the explicit variations
\begin{equation*}
  \delta M^2 = (m_L^2 - m_T^2)\,\delta P_L \;,\quad \delta
  I^{(1)}=(I_L^{(1)}-I_T^{(1)})\,\delta P_L = (m_L^2-m_T^2)\,\tilde
  I^{(2)}_{TL}(0)\,\delta P_L
\end{equation*}
we can rewrite the second equation in eqs.~(\ref{eq:sym-cond-1.3}) as
\begin{equation}
  \label{eq:par-rel-2}
  m^2_L-m^2_T= \tfrac12\,v^2\,\tilde \theta_4 (0)
\end{equation}
In the second step, we impose invariance of the background field equation
(\ref{eq:gen-feat1-on}) by requiring
\begin{equation}
  \label{eq:sym-cond-2.1}
  \delta \F'=0   
\end{equation}
This can be written, by using eqs.~(\ref{eq:gen-feat2-on})
and~(\ref{eq:sym-cond-1.1}) and the independence of $\xi$ and $u$, as
\begin{equation*}
  \delta \Omega^{(2)}=\tfrac12\, \delta \theta \quad,\qquad \delta
  \Omega^{(1)}-\delta M^2 -
  (\Omega^{(2)}-\tfrac12\,\theta)\,v^2\,\delta P_L
\end{equation*}
Explicit tensorial calculation for the first equation yields
\begin{equation}
  \label{eq:par-rel-3}
  \begin{split}
    &\Omega^{(2)}_2 =\Omega^{(2)}_4 + \tfrac12 [ \theta_2 -
      \theta_4]\\ &\Omega^{(2)}_3= \Omega^{(2)}_1 - \Omega^{(2)}_4 +
    \tfrac12 [ \theta_3 - \theta_1 + \theta_4\,] \\ & \Omega^{(2)}_5 =
    \Omega^{(2)}_1 - \tilde \Omega^{(2)}_4 + \tfrac12 [\, \theta_5 -
      \theta_1 + \theta_4\,]
  \end{split}
\end{equation}
While for the second we obtain
\begin{equation}
  \label{eq:par-rel-4}
  \Omega^{(1)}=- \tilde \Omega^{(2)}_4 (0) \,v^2\,P_T
\end{equation}
In conclusion, according to
eqs.~(\ref{eq:par-rel-1}),~(\ref{eq:par-rel-2}),~(\ref{eq:par-rel-3})
and~(\ref{eq:par-rel-4}), the actual independent parameters are
$\theta_1$, $\theta_4$, $m_L^2$, $\Omega^{(2)}_1$ and
$\Omega^{(2)}_4$. The integrated versions of eqs.~(\ref{eq:sym-cond-1.1})
and~(\ref{eq:sym-cond-2.1}) are obtained by writing $M^2+\V$ and
$\F'$ in a manifestly $O(N)$ symmetric form
\begin{equation*}
  \begin{split} 
    &M^2 + \V = \mu^{(1)} {\bf 1} + \tfrac12 \, \gamma^{(1)} [ \xi +
      I]\;,\quad \F'= \mu^{(2)}\,{\bf 1} + \tfrac12 \,
    \gamma^{(2)}\,\xi + \tfrac12 \, [ M^2 + \V ]\\[1.8mm]
    &\gamma^{(\alpha)}_{ijkm}= \gamma^{(\alpha)}_1 \,\delta_{ij}
    \delta_{km} + \tfrac12 \,\gamma^{(\alpha)}_2\,[\delta_{ik}
      \delta_{jm} + \delta_{im} \delta_{jk}]\;,\quad \alpha=1,2
  \end{split}
\end{equation*}
with $\gamma^{(\alpha)}$ and $\mu^{(\alpha)}$ defined in terms of the
free parameters as follows
\begin{equation}
\label{eq:symm-form}
 \begin{split}
   &\gamma^{(1)}_1=[\theta_1 - \theta_4
   -\tfrac12 \,\theta_4 \,\theta_1\,(I_{LL}^{(2)} -
   I_{TL}^{(2)})]\,[{\bf 1} + \tfrac12\,\theta_4\,( I_{TL}^{(2)} -
   I_{TT}^{(2)})][{\bf 1} - \tfrac12\, \theta_4\,
   I_{TL}^{(2)}]^{-1}\,\Delta^{-1}\\[2mm]
   &\Delta= [{\bf 1} +
   \tfrac12\,\theta_4\,[\,I_{TL}^{(2)} - N I_{TT}^{(2)}]][{\bf 1} +
   \tfrac12\,\theta_1\,I_{LL}^{(2)}] -
   \tfrac{N-1}{2}\,\theta_1\,I_{TT}^{(2)}[{\bf 1} + \tfrac12\theta_4
   \,I_{TL}^{(2)}]^{-1}\\[2mm]
   &\gamma^{(1)}_2= \theta_4\,[{\bf 1}+ \tfrac12\,
   I_{TL}^{(2)}]^{-1} \\[2mm] 
   &\gamma^{(2)}_1= \tfrac12\,\Omega^{(2)}_3
   - \tfrac14\, \theta_3 \quad , \qquad
   \gamma^{(2)}_2= \tfrac12\,\Omega^{(2)}_4 - \tfrac14\,\theta_4
   \\[2mm] &\mu^{(1)}= m_L^2 - \tfrac12 \,
   \tilde\gamma^{(1)}_2(0)\, [v^2 + I^{(1)}_L]- \tfrac12 \,
   \tilde\gamma^{(1)}_1(0)\, [v^2 + I^{(1)}_{L}+
   (N-1)\,I^{(1)}_T]\\[2mm] &\mu^{(2)}= - \tfrac12 \, m_L^2  
   - \tfrac12\,\tilde\Omega^{(2)}_1(0)\,v^2 + \tfrac14\,
   \tilde\theta_1(0)\, v^2
 \end{split}
\end{equation}
One can verify that the standard HF definitions of the tunable parameters
indeed satisfy
eqs.~(\ref{eq:par-rel-1}),~(\ref{eq:par-rel-2}),~(\ref{eq:par-rel-3})
and~(\ref{eq:par-rel-4}) and that the corresponding parameters of the
manifestly $O(N)$ symmetric form are
\begin{equation*}
   \gamma^{(1)}_2=2\,\gamma^{(1)}_1=4\,\gamma^{(2)}_2=\tfrac23 \,
   \lambda_0 \, {\bf 1} \; , \quad \gamma^{(2)}_1=0\;,\quad
   \mu^{(1)}= \tfrac12 \, m^2_0 \; , \quad \mu^{(2)}= 0
\end{equation*}
Now, as in sec.~\ref{sec:defin-modif-hf}, we require that $\theta$ and
$\Omega^{(2)}$ have the same general leading log structure characteristic
of their HF definitions [see eqs.~(\ref{eq:HF-theta-comps}) and
eqs.~(\ref{eq:HF-Om2-comps})], namely
\begin{equation}
\label{eq:eq:genLL-1-on}
  \begin{split}
    &\Omega^{(2)}_1=\lambda_0\,F_1
    (\lambda_0I_{LL}^{(2)},\,\lambda_0I_{TL}^{(2)},\,\lambda_0\,I_{TT}^{(2)}
    ;\,N)
    \;,\quad \theta_1=\lambda_0\,F_2
    (\lambda_0I_{LL}^{(2)},\,\lambda_0I_{TL}^{(2)},\,\lambda_0\,I_{TT}^{(2)}
    ;\,N)\\
    &\Omega^{(2)}_4=\lambda_0\,F_3
    (\lambda_0I_{LL}^{(2)},\,\lambda_0I_{TL}^{(2)},\,\lambda_0\,I_{TT}^{(2)}
    ;\,N)
    \;,\quad \theta_4=\lambda_0\,F_4
    (\lambda_0I_{LL}^{(2)},\,\lambda_0I_{TL}^{(2)},\,\lambda_0\,I_{TT}^{(2)}
    ;\,N)\\
  \end{split}
\end{equation}
where the $F_A(x,y,z;N)$ are generic functions of commuting arguments since
the operators $I_{LL}^{(2)}$, $I_{TL}^{(2)}$ and $I_{TT}^{(2)}$ are
diagonal in Fourier space.

Notice that the same structure is inherited by all the others
components of $\theta$ and $\Omega^{(2)}$ as can be verified by
eqs.~(\ref{eq:par-rel-1}) and eqs.~(\ref{eq:par-rel-3}). We assume a
similar structure also for $m_L^2$ as we did in
eq.~(\ref{eq:genLL-2-n=1}) for the $N=1$ case
\begin{equation}
  \label{eq:eq:genLL-2-on}
    m^2_L= \lambda_0\,v^2\, F_5
    (\lambda_0 \tilde I_{LL}^{(2)}(0),\,\lambda_0\,\tilde I_{TL}^{(2)}(0),
    \lambda_0\tilde I_{TT}^{(2)}(0);\,N)
\end{equation}
Notice that, by means of eq.~(\ref{eq:par-rel-2}), the same structure
is inherited by $m^2_T$ which is consistent with its HF
definition [see eqs.~(\ref{eq:HF-pardef-on})]. 

We have thus reduced our parameterization freedom to five functions $F_A$
of three variables. We still have to require that these functions fulfill
some further properties that hold true in the HF approximation.

Hartree--Fock matches perturbatively at 1--loop order. This matching
requires that $\theta$ and $m_L^2$ match at tree level and that
$\Omega^{(2)}$ matches at tree level and 1--loop order. Actually
$\Omega^{(2)}_4$ must match only at 1-loop level since the tree level
term never appears in the vertex functions. Explicitly we have the
following conditions on the $F$'s
\begin{equation*}
  \begin{split}
    &F_1(x,y,z;N)= \tfrac16+ \tfrac14\,x +\tfrac1{36}(N-1)\,z + \ldots
    \quad,\qquad F_2(x,y,z;N)= 1+ \ldots\\ &F_3(x,y,z;N)=c_0+\tfrac19\,y
    + \ldots \quad \qquad \qquad \qquad,\qquad F_4(x,y,z;N)=
    \tfrac23+ \ldots\\ &F_5(x,y,z;N)=\tfrac13+\ldots
  \end{split}
\end{equation*}
where $c_0$ is a purely numerical arbitrary constant. One can easily check
that the above conditions, together with the symmetry relations in
eqs.~(\ref{eq:par-rel-1}),~(\ref{eq:par-rel-2}),~(\ref{eq:par-rel-3})
and~(\ref{eq:par-rel-4}), are enough to guarantee the matching at 1--loop
order of all the vertex functions.

In the $N \to \infty$ limit the HF resummation reproduces
correctly the equations of the usual large $N$ approximation. This
can be done as follows. First we restrict to the special
case of a background field which maintains a fixed direction (i.e. the
direction of the vacuum expectation value $v$) , that is $\phi_i(x) =
\phi (x)\, \hat{v}_i$. Then we can reduce the equations of motion into
a projected form by setting
\begin{equation*}
    u_{\bm{k}}= u_{\bm{k}\,L}\,P_L + u_{\bm{k}\,T}\,P_T \quad,\qquad
    \V= \V_L\,P_L + \V_T\,P_T
\end{equation*}             
By substituting into
eqs.~(\ref{eq:gen-feat1-on}),~(\ref{eq:gen-feat2-on}),~(\ref{eq:gen-mfeq-on})
and projecting one obtains
\begin{equation*}
  \begin{split} 
    &[\Box + \V_L + (\Omega^{(2)}_1 - \tfrac12\,\theta_1)\Delta\,\xi]
    \phi=0\\[2mm] &[\Box + m_T^2+\V_T] u_{\bm{k}\,T} =0 \;,\quad 
    ~~~~~~~~~~~~~~~~~~[\Box +m_L^2 + \V_L] u_{\bm{k}\,L}=0 \\[2mm] 
      &\V_L = \tfrac12\,\theta_1\,[
      \Delta \xi + J_L ] + \tfrac12(N-1)\,\theta_5\, J_T\;,\quad \V_T =
    \tfrac12\,\theta_5\, [\Delta \xi + J_L]+
                    [\tfrac12(N-1)\,\theta_3+ \tfrac12\,\theta_2]\,J_T
  \end{split}
\end{equation*}  
where $\xi=\phi^2$ and $J_L$ and $J_T$ have the obvious meaning [see
eq.~(\ref{eq:J-def-on})].  Now let us rescale the coupling, the
background field and the vacuum expectation value as prescribed by the
standard large $N$ procedure
\begin{equation*}
  \lambda_0 \to \lambda_0/N \qquad \phi (x) \to \sqrt{N}\,\phi (x)
  \qquad v^2 \to N \, v^2
\end{equation*}
By taking the limit and using the symmetry conditions one can
see that the correct large $N$ equations
\begin{equation}
  \label{eq:large-N-eom}
  \begin{split} 
    &[\Box + \V_T] \phi=0 \quad,\qquad [\Box + \V_T] u_{\bm{k}\,T}
    =0\\[2mm] & \V_T = \tfrac16 \, \lambda \, \Delta \xi + \tfrac16
    \, \lambda \, \int_{|\bm p|<\Lambda} \frac{d^3 p}{(2 \pi)^3}
    \left[ u_{\bm p}(x)\, u_{\bm p}^\dag(y) - \frac{1}{2\,
        \omega_{{\bm p}, T}} \right]\\[2mm] &\lambda^{-1} =
    \lambda_0^{-1}- \tfrac16 \, \tilde I_{TT}^{(2)}({q_s})
  \end{split}
\end{equation}  
are recovered, provided the limits of the free parameters satisfy the
following relations
\begin{equation}
  \label{eq:largeN-match-cons}
  \begin{split} 
      &\tfrac12\,v^2\,\tilde \theta_4 (0) - m_L^2 \to 0\\[2mm]
      &N\,(\theta_1 - \theta_4) \to \tfrac13 \,\lambda_0\,[{\bf 1} -
        \tfrac16 \,\lambda_0 \,I_{TT}^{(2)}]^{-1}\\[2mm] &N\,
      \Omega^{(2)}_1 \to \tfrac16\lambda_0\,[{\bf 1} - \tfrac16 \, \lambda_0
        \,\tilde I_{TT}^{(2)}]^{-1}
    \end{split} 
\end{equation}  
Suitable conditions on the functions $F_A$ then follow from
eqs.~(\ref{eq:eq:genLL-1-on}). Notice that in the integrals $I_{TT}^{(2)}$
in eqs.~(\ref{eq:large-N-eom}) and~(\ref{eq:largeN-match-cons}) we have
$m^2_T=0$, as consistent with the limits of the parameters.

To conclude, in the HF approximation $\Omega^{(2)}_1$ and the
$\theta_\alpha$ fulfill positivity conditions. More precisely they are real
and positive when evaluated at the purely spatial value of the momentum $q_s$
[see eq.~(\ref{eq:simm-point})].

We now proceed in defining our modified HF approximation by making some
further sensible requirements that are not satisfied by the HF
approximation.

\vskip15pt
First of all we require our approximation to be gapless. That is to
say we require that the transverse mass of the internal propagator is
zero. As a consequence by eq.~(\ref{eq:par-rel-2}) we have
\begin{equation}
\label{eq:gapless-cond}
  m^2_L=\tfrac12\,v^2\,\theta_4(0)\Longrightarrow
  F_5 (x,y,z;N) = \tfrac12\,F_4 (x,y,z;N)
\end{equation}
Due to this condition the transverse internal propagators are now
massless and therefore the Goldstone bosons loop integral
$I_{TT}^{(2)}$ is logarithmically IR--divergent. The symmetry
relations in eq.~(\ref{eq:gapless-cond}) and eq.~(\ref{eq:par-rel-4})
define the longitudinal mass $m^2_L$ and $\Omega^{(1)}$ in terms of
the zero momentum values of $\theta_4$ and $\Omega_4^{(2)}$
respectively. To avoid IR divergences we then require
that $\theta_4$ and $\Omega_4^{(2)}$ do not depend at all on
$I_{TT}^{(2)}$. Thus we can write
\begin{equation*}
  F_3(x,y,z;N)=K_3(x,y;N)\quad , \qquad F_4(x,y,z;N)=K_4(x,y;N)
\end{equation*}
where $K_A(x,y;N)$ ($A=3,4$) are arbitrary functions of two variables only
(plus $N$). The same argument apply to the parameters $\mu^{(1)}$ and
$\mu^{(2)}$ of the manifestly symmetric form in
eq.~(\ref{eq:symm-form}): we require that $\tfrac12\,\theta_1-
\Omega_1^{(2)}$ and $\gamma_2^{(2)}$ do not depend on
$I_{TT}^{(2)}$, which in turn implies
\begin{equation*}
  \begin{split}
    &F_2(x,y,z;N)=\frac{K_4(x,y;N)}{1 + \tfrac12 \, K_4(x,y;N) (y-x)} -
    \frac{K_4(x,y;N)}{(N-1)\frac{[1 + \tfrac12 \, K_4(x,y;N) (y-x)]^2}{1
        + \tfrac12 \, K_4(x,y;N) (y-z)}+K_2(x,y;N)}\\ &F_1(x,y,z;N)= -
    \frac{1}{2} \frac{K_4(x,y;N)}{(N-1)\frac{[1 + \tfrac12 \, K_4(x,y;N)
          (y-x)]^2}{1 + \tfrac12 \, K_4(x,y;N) (y-z)}+ K_2(x,y;N)}+K_1(x,y;N)
  \end{split}
\end{equation*}
where, again, $K_A(x,y;N)$ ($A=1,2$) are arbitrary functions of two
variables (and $N$).

Now, as in sec.~\ref{sec:defin-modif-hf}, we require the
renormalizability with the 1--loop beta function. That is, we assume
the following RG equation for the bare coupling $\lambda_0$ 
\begin{equation*}
  \frac{\partial \lambda_0}{\partial \log{\Lambda}} = \frac{N+8}{24
    \pi^2}\lambda_0^2 + O(\Lambda^{-1})
\end{equation*}
and ask that the free parameter function $K_A$ do not depend on
$\log{\Lambda}$. By the same procedure of
sec.~\ref{sec:defin-modif-hf} and using
\begin{equation*} 
  \frac{\partial\,I_{LL}^{(2)} }{\partial \log{\Lambda}}=
  \frac{\partial I_{TT}^{(2)}}{\partial \log{\Lambda}}+ O(\Lambda^{-1}) 
  =\frac{\partial I_{TL}^{(2)}}{\partial\log{\Lambda}}+ O(\Lambda^{-1}) 
  = -\frac1{8 \pi^2}\,{\bf 1} + O(\Lambda^{-1})
\end{equation*}
we obtain the general forms
\begin{equation*}
  \begin{split}  
    &K_j(x,y;N)= \frac{f_j (\alpha(x,y);N)}{1 - \tfrac16(N+8)\,x}
    \quad , \qquad j=1,3,4 \\ &K_2(x,y;N)= f_2 (\alpha(x,y);N)
   \end{split}
\end{equation*}
where 
\begin{equation*}
  \alpha(x,y)= \frac{x-y}{1 - \tfrac16(N+8) \, x}
\end{equation*}
Recalling that $x=\lambda_0I_{LL}^{(2)}$ and $y=\lambda_0I_{TL}^{(2)}$
we see that the Fourier transform of $\alpha$ positive definite for
purely spatial value of the momentum.  Notice also that $\alpha$
vanishes when $v \to 0$ since $I_{LL}^{(2)}$ and $I_{TL}^{(2)}$
coincide in this limit.

Notice that now, in contrast to the $N=1$ case studied in
sec.~\ref{sec:defin-modif-hf} [see eqs.~(\ref{eq:mHF-pardef-n=1})], the
requirement of renormalizability does not fix completely the form of the
free parameters. This occurs because for $N>1$ there exist three distinct
finite parts associated with the logarithmic cutoff divergence of
$I_{LL}^{(2)}$, $I_{TL}^{(2)}$ and $I_{TT}^{(2)}$.  So now we are left
with four arbitrary functions $f_A$ of one variable (and $N$) to determine.

It is convenient to rewrite all the constraints previously found in terms
of this new parametrization. After some straightforward, albeit rather
involved calculations, one obtains
\begin{itemize}
\item Perturbative matching:
  \begin{equation*}
    \begin{split}
      &f_1(\alpha;N)= \tfrac1{36}\left[(N-1)-3\, f'_2(0;N) - 9\,
      f'_4(0;N)\right]\alpha + O(\alpha^2)\\[2mm] &f_4(\alpha;N)= 2/3
      + O(\alpha)\;,\quad f_3(\alpha;N)= \frac{2}{3(N+8)} -
      \tfrac1{9}\, \alpha + O(\alpha^2)\\[2mm] &f_2(\alpha;N)= -(N+1)
      + O(\alpha)
    \end{split}
  \end{equation*}  
\item Matching in the large--$N$ limit:
  \begin{equation*}
    f_1(\alpha/ N;N) \to 0 \;,\quad f_4(\alpha/ N;N) \to f(
    \alpha)\;,\quad f_2(\alpha/ N;N)+ N \to 1 - \tfrac{6-\alpha}2 \,
    f(\alpha)
  \end{equation*}
  where the arbitrary function $f$ enters only the decoupled longitudinal
  sector.
\item Positivity constraints on $\Omega^{(2)}_1$ and $\theta_\alpha$:
  on the real interval $0\le \zeta \le 6/(N+8)$
  \begin{equation*}
    0<f_4(\zeta;N)<\frac{2}{\zeta} \;, \quad f_2(\zeta;N)+N-1<0
    \;,\quad f_1(\zeta;N)>0 
  \end{equation*}              
  Actually these conditions are slightly stronger than strictly
  necessary. In any case they also guarantee the positivity of all the
  $\theta_\alpha$ parameters (which holds true in the usual HF
  approximation as well).
\end{itemize}
One last requirement is that the $N=1$ case should be recovered for all the
parameters that have meaning also in this case. These are $\Omega^{(2)}_1$,
$\theta_1$ and $\theta_4$ (that, evaluated at zero momentum, determines the
longitudinal mass according to eq.~(\ref{eq:gapless-cond})). When $N=1$
they should have the following form [see eqs.~(\ref{eq:mHF-pardef-n=1})]
\begin{equation*}
  \theta_1=12\, \Omega^{(2)}_1=\tfrac32 \, \theta_4=\lambda_0\,[{\bf
      1} - \tfrac32 \, \lambda_0 \,I_{LL}^{(2)}]^{-1}
\end{equation*}
that is, in terms of the functions $f$'s,
\begin{equation*}
  \begin{split}  
    f_4 (\alpha;1)=\tfrac23 \quad,\quad f_2(\alpha;1)=- \frac{2 -
      \tfrac23 \alpha}{1 - \alpha}\quad,\quad f_1(\alpha;1)=
   \frac{\alpha}{9 - 3 \alpha}
  \end{split}
\end{equation*}
Notice also that all the possible choices of free parameters coincide
when $v\to 0$, since $\alpha$ vanishes in this limit and the
perturbative constraints uniquely fix the form of the $f$'s when
$\alpha=0$. Moreover one can verify that the resulting $v=0$ improved
HF coincide with the massless limit of the improved approximation
in the unbroken symmetry phase defined in ref.~\cite{Destri:2005qm}.

In conclusion all possible forms of the functions $f$'s that
fulfill the above matching constraints define improved Hartree--Fock
resummations with the required features of gaplessness and
renormalizability.  One simple choice for the $f$'s is
\begin{equation}
  \label{eq:mHF-ex-1}
  \begin{split} 
    &f_4(\alpha;N)=2/3 \quad,\qquad f_2(\alpha;N)=-(N+1)\,
    \frac{1 - \tfrac13 \alpha}{1 - \alpha}+(N-1)\,\alpha\\
    &f_1(\alpha;N)=  \frac{\alpha}{9 - 2 \alpha} \quad,\qquad
    f_3(\alpha;N)=\frac{2}{3\,(N+8)}\,\frac{1}{1+\tfrac{N+8}6\,\alpha}
  \end{split} 
\end{equation}
As already remarked, this choice is not
unique. For example we can consider a second form 
\begin{equation}
  \label{eq:mHF-ex-2}
  \begin{split} 
    &f_4(\alpha;N)= \frac{2/3}{1 + \tfrac{N-1}6 \,
    \alpha}\;,\quad f_2(\alpha;N)=-(N+1)\, \frac{1 - \tfrac13
    \alpha}{1 - \alpha}+\tfrac{4(N-1)}3\,\alpha\,\,\frac{1 +
    \tfrac{N+8}{12} \,\alpha}{1 + \tfrac{N+8}6 \,\alpha}\\
    &f_1(\alpha;N)=  \frac{\alpha}{9 - 2 \alpha}\quad,\qquad
    f_3(\alpha;N)=\frac{2}{3\,(N+8)}\,\frac{1}{1+\tfrac{N+8}6\,\alpha}
  \end{split} 
\end{equation}
As already explained in sec.~\ref{sec:defin-modif-hf}, our modified
resummation adds leading logarithm contributions of diagrams that are
not present in the usual HF approximation; then the two forms just
provided of the free parameters, as well as all the other possible
ones, correspond to different choices of the associated finite parts.

\vskip15pt 

Given one specific choice for the free parameters, we can proceed in
applying the coupling constant renormalization condition in
eq.~(\ref{eq:recon-coumod-on}). Notice that the consistence of this
renormalization requires that $\Omega^{(2)}_1$ is monotonically
growing with $\lambda_0$ for any given purely spatial momentum. We
have omitted to include this requirement in the previous general
discussion since it would lead in general to rather complicated
constraints. It holds true for the two simple examples given in
eqs.~(\ref{eq:mHF-ex-1}) and eqs.~(\ref{eq:mHF-ex-2}), as one can
explicitly check. Moreover it holds true in general (i.e. for any
improved HF resummation) for scales such that $s^2\gg v^2$, since we
have in this limit
\begin{equation*}
  \tilde \Omega^{(2)}_1(q_s)=  \frac{\tfrac16\lambda_0}{1-
    \tfrac16 (N+8)\, \lambda_0 \,\tilde I^{(2)}(q_s)}+
  O(m^2/s^2)
\end{equation*} 
where $I^{(2)}$ stands for anyone of $I_{LL}^{(2)}$, $I_{TL}^{(2)}$ and
$I_{TT}^{(2)}$. The renormalization condition thus defines the
bare--to--renormalized relation
\begin{equation*}
  \lambda = \frac{\lambda_0}{1- \tfrac{N+8}6\, \lambda_0
   \, \tilde I^{(2)}(q_s)}+ O(m^2/s^2)
\end{equation*} 
which is the usual 1 loop RG--invariant relation up to $O(m^2/s^2)$
terms. This shows that the direct coupling renormalization condition
is approximately scale invariant for high renormalization scales. To
obtain complete scale invariance it is enough to slightly modify the
bare--to--renormalized parameterization by changing
the renormalization condition in the following way
\begin{equation}
  \label{eq:mHF-coren-mod}
  \begin{split}
    6\,\tilde \Omega^{(2)}_1(q_s) &= 6\,\lambda_0\,F_1
    (\lambda_0\tilde I_{LL}^{(2)}(q_s),\,\lambda_0\tilde I_{TL}^{(2)}(q_s),\,
    \lambda_0\,\tilde I_{TT}^{(2)}(q_s);\,N)\\[1.5mm]
    &\substack{=\\{\rm def}}\; 6\,\lambda \, F_1(\lambda\tilde
    J_{LL}^{(2)}(q_s),\,\lambda \tilde J_{TL}^{(2)}(q_s),\, \lambda \tilde
    J_{TT}^{(2)}(q_s);\,N)\\[1.5mm]
    &=\lambda + O(m^2/s^2)
  \end{split}
\end{equation}
where 
\begin{equation*}
  \tilde J_{LL}^{(2)}(p) = I_{LL}^{(2)}-\tilde I_{LL}^{(2)}(q_s){\bf
    1} \;,\quad\tilde J_{TL}^{(2)}(p) = I_{TL}^{(2)}-\tilde
  I_{LL}^{(2)}(q_s){\bf 1} \;,\quad\tilde J_{TT}^{(2)}(p) =
  I_{TT}^{(2)}-\tilde I_{LL}^{(2)}(q_s){\bf 1}
\end{equation*}
are the properly subtracted finite loops (notice that the subtraction term
is always the purely longitudinal $I_{LL}^{(2)}(q_s)$). Then, using the
parameterization of $F_1$ in terms of the $f_i(\alpha;N)$, one can verify
that the bare--to--renormalized relation reads exactly
\begin{equation}
  \label{eq:mHF-btor-mod}
  \lambda = \frac{\lambda_0}{1- \tfrac{N+8}6\, \lambda_0\,
    \tilde I_{LL}^{(2)}(q_s)}
\end{equation} 
Comparing this with eq.~(\ref{eq:1LRGI-btorrel-on}) we can see that this
has the correct $1-$loop-RG improved behaviour with a specific choice of
the finite parts. Then all the free parameters take the following
manifestly finite forms
\begin{equation*}
  \begin{split}
    &\Omega^{(2)}_1=\lambda \,F_1(\lambda J_{LL}^{(2)}, \,\lambda 
    J_{TL}^{(2)},  \,\lambda J_{TT}^{(2)}; \,N )\quad,\qquad 
    \theta_1 =\lambda \, F_2(\lambda J_{LL}^{(2)},\,\lambda
    J_{TL}^{(2)},\,\lambda J_{TT}^{(2)};\,N )\\ 
    &\Omega^{(2)}_4 =\lambda \, F_4(\lambda \,
    J_{LL}^{(2)},\lambda \, J_{TL}^{(2)}, \lambda \, J_{TT}^{(2)};N
    )\quad,\qquad\theta_4 =\lambda \, F_4(\lambda J_{LL}^{(2)},\,\lambda 
    J_{TL}^{(2)}, \,\lambda J_{TT}^{(2)};\,N )
  \end{split}
\end{equation*}
Moreover, consistently with the renormalization condition in
eq.~(\ref{eq:mHF-coren-mod}), we can define the running coupling
constant by means of the following equation
\begin{equation*}
   \, \lambda(p) \, F_1(\lambda(p) \tilde J_{LL}^{(2)}(p),\,\lambda(p) 
   \tilde J_{TL}^{(2)}(p), \,\lambda(p) \tilde J_{TT}^{(2)}(p);\,N)= 
   \tilde \Omega^{(2)}_1(p) 
\end{equation*}
whose solution is simply the extension of eq.~(\ref{eq:mHF-btor-mod}) to
arbitrary momentum
\begin{equation*}
  \lambda(p)= \frac{\lambda}{1- \tfrac{N+8}6\,\lambda\,
    \tilde J_{LL}^{(2)}(p)}
\end{equation*} 
In term of this and of the subtracted integrals we can write the
renormalized form of the Fourier transform of $\alpha$:
\begin{equation*}
  \tilde \alpha (p) = \lambda (p)\,[\,\tilde  J_{TL}^{(2)}(p) -\tilde
  J_{LL}^{(2)}(p) \,]
\end{equation*}
Notice that $\alpha$ becomes small for large (but smaller than the
Landau pole) spatial momentum $q_s$. More precisely it is of order
$O(\lambda(q_s)\,v^2/s^2)$. Because of the Landau pole, the cut--off
of the theory cannot be removed but should be fixed to values suitably
smaller than the pole (i.e such that $\lambda(q_\Lambda)$ is of order
one). The mass scale of the theory, that is $v$, is much smaller than
the cut--off itself. Therefore for spatial momenta with values near the
cut--off $\alpha$ is small. Then the perturbative (in $\alpha$)
matching conditions assure that the different allowed choices of the
free parameters give the same results to order
$O(\lambda(q_\Lambda)\,v^2/\Lambda^2)$. In this sense we can say that
all the class of improved HF approximations shares the same UV
behaviour.

One can easily check that if we renormalize at scale $s$ with
coupling constant $\lambda$ and at scale $s'$ with coupling constant
$\lambda'=\lambda (q_{s'})$ we define the same bare coupling
constant. That is to say that the bare--to--renormalized relation in
eq.~(\ref{eq:mHF-btor-mod}) is RG--invariant. As a consequence the
expressions of the parameters in terms of $\lambda_0$ and $I^{(2)}$ in
eqs.~(\ref{eq:eq:genLL-1-on}) can be thought as manifestly scale
invariant definitions.

\section{Conclusions}
In this paper we extended to the case of spontaneously broken symmetry the
improvement of the HF approach to the $O(N)$ scalar theory that was
proposed in ref.~\cite{Destri:2005qm}. In contrast to the standard HF
approximation, our improved one is renormalizable, RG--invariant and
correctly gapless in the Goldstone sector. However, it is not unique,
except for $N=1$, because the mass difference between the longitudinal
sector and the transverse Goldstone sector allows for a richer structure of
renormalization finite parts that cannot be fully restricted by the
requirements of renormalizability and RG--invariance. As a consequence, an
entire class of improvements is identified, parametrized by the four
functions $f_A(\alpha;\,N)$ which must satisfy some further constraints
explicitly written in the previous section. It is important to stress that,
albeit of functional type, the remaining freedom is much smaller than what
simple dimensional analysis would allow even when simple
momentum--independent observables are concerned (the momentum dependences
in our approximation are anyway fixed by construction to be of mean--field
type). This makes it difficult, although not a priori impossible, to
identify consistent schemes to further constrain the functions
$f_A(\alpha;\,N)$ by requiring agreement in suitable calculations beyond
one loop. Work on this direction is in progress.


\begin{thebibliography}{100}

\bibitem{Berges:intro}
  J.~Berges, ``Introduction to nonequilibrium quantum field theory,''
  AIP Conf.\ Proc.\  {\bf 739} (2005) 3
  [arXiv:hep-ph/0409233].

  J.~Berges and J.~Serreau,
  ``Progress in nonequilibrium quantum field theory. II,''
  arXiv:hep-ph/0410330.

  J.~Berges and J.~Serreau,
  ``Progress in nonequilibrium quantum field theory,''
  arXiv:hep-ph/0302210.

\bibitem{Manfredini:2000sk}
  E.~Manfredini,
  ``Aspects of non--equilibrium dynamics in quantum field theory,''
  arXiv:hep-ph/0101202.

\bibitem{Cooper:largeN}
  F.~Cooper and E.~Mottola,
  ``Initial Value Problems In Quantum Field Theory In The Large N
  Approximation,''
  Phys.\ Rev.\ D {\bf 36}, 3114 (1987).

  F.~Cooper, S.~Habib, Y.~Kluger, E.~Mottola, J.~P.~Paz and P.~R.~Anderson,
  ``Nonequilibrium quantum fields in the large N expansion,''
  Phys.\ Rev.\ D {\bf 50} (1994) 2848
  [arXiv:hep-ph/9405352].

\bibitem{Boyanovsky:1994me}
  D.~Boyanovsky, H.~J.~de Vega, R.~Holman, D.~S.~Lee and A.~Singh,
  ``Dissipation via particle production in scalar field theories,''
  Phys.\ Rev.\ D {\bf 51}, 4419 (1995)
  [arXiv:hep-ph/9408214].

\bibitem{Baacke:largeN}
  J.~Baacke and K.~Heitmann, ``Nonequilibrium evolution and symmetry
  structure of the large-N $\phi^4$ model at finite temperature'',
  Phys.\ Rev.\ D {\bf 62}, 105022 (2000) [arXiv:hep-ph/0003317].

  J.~Baacke, K.~Heitmann and C.~Patzold,
  ``Renormalization of nonequilibrium dynamics at large N and finite
  temperature,''
  Phys.\ Rev.\ D {\bf 57} (1998) 6406
  [arXiv:hep-ph/9712506].

\bibitem{Destri:largeN}
  C.~Destri and E.~Manfredini,
  ``Out-of-equilibrium dynamics of large-N $\phi^4$ QFT in finite volume,''
  Phys.\ Rev.\ D {\bf 62} (2000) 025007
  [arXiv:hep-ph/0001177].

\bibitem{Amelino-Camelia:1992nc}
  G.~Amelino-Camelia and S.~Y.~Pi,
  ``Selfconsistent improvement of the finite temperature effective potential,''
  Phys.\ Rev.\ D {\bf 47} (1993) 2356
  [arXiv:hep-ph/9211211].

\bibitem{Destri:1999he}
  C.~Destri and E.~Manfredini,
  ``An improved time-dependent Hartree-Fock approach for scalar $\phi^4$ QFT,''
  Phys.\ Rev.\ D {\bf 62} (2000) 025008
  [arXiv:hep-ph/0001178].

\bibitem{Michalski}
S.~Michalski,
``Nonequilibrium dynamics of the O(N) linear sigma model in the Hartree
approximation,''
arXiv:hep-ph/0301134.

\bibitem{Matsui}
T.~Matsui,
``Variational approach to dynamics of quantum fields,''
arXiv:hep-ph/0111277.

Y.~Tsue, D.~Vautherin and T.~Matsui,
``Mean field theory for collective motion of quantum meson fields,''
Prog.\ Theor.\ Phys.\  {\bf 102}, 313 (1999)
[arXiv:hep-ph/9812254].

\bibitem{Berges:2000ur}
J.~Berges and J.~Cox,
``Thermalization of quantum fields from time-reversal invariant evolution
equations,''
Phys.\ Lett.\ B {\bf 517} (2001) 369
[arXiv:hep-ph/0006160].

\bibitem{Habib:irrev}
  S.~Habib, Y.~Kluger, E.~Mottola and J.~P.~Paz,
  ``Dissipation and Decoherence in Mean Field Theory,''
  Phys.\ Rev.\ Lett.\  {\bf 76} (1996) 4660
  [arXiv:hep-ph/9509413].

  F.~Cooper, S.~Habib, Y.~Kluger and E.~Mottola,
  ``Nonequilibrium dynamics of symmetry breaking in lambda $\phi^4$ field
  theory,''
  Phys.\ Rev.\ D {\bf 55}, 6471 (1997)
  [arXiv:hep-ph/9610345].

B.~Mihaila, T.~Athan, F.~Cooper, J.~Dawson and S.~Habib,
``Exact and approximate dynamics of the quantum mechanical O(N) model,''
Phys.\ Rev.\ D {\bf 62}, 125015 (2000)
[arXiv:hep-ph/0003105].

\bibitem{Boyanovsky:meanfield}
  D.~Boyanovsky and H.~J.~de Vega,
  ``Quantum rolling down out-of-equilibrium,''
  Phys.\ Rev.\ D {\bf 47} (1993) 2343
  [arXiv:hep-th/9211044].

  D.~Boyanovsky, H.~J.~de Vega, R.~Holman and J.~F.~J.~Salgado,
  ``Analytic and numerical study of preheating dynamics,''
  Phys.\ Rev.\ D {\bf 54}, 7570 (1996)
  [arXiv:hep-ph/9608205].

  D.~Boyanovsky, C.~Destri, H.~J.~de Vega, R.~Holman and J.~Salgado,
  ``Asymptotic dynamics in scalar field theory: Anomalous relaxation,''
  Phys.\ Rev.\ D {\bf 57}, 7388 (1998)
  [arXiv:hep-ph/9711384].

\bibitem{Wetterich:meanfield}
C.~Wetterich,
``Nonequilibrium time evolution in quantum field theory,''
Phys.\ Rev.\ E {\bf 56} (1997) 2687
[arXiv:hep-th/9703006].
 
L.~M.~A.~Bettencourt and C.~Wetterich,
``Time evolution of correlation functions in non--equilibrium field
theories,''
Phys.\ Lett.\ B {\bf 430} (1998) 140
[arXiv:hep-ph/9712429].

  G.~F.~Bonini and C.~Wetterich,
  ``Time evolution of correlation functions and thermalization,''
  Phys.\ Rev.\ D {\bf 60}, 105026 (1999)
  [arXiv:hep-ph/9907533].

\bibitem{Smit:meanfield}
M.~Salle and J.~Smit,
``The Hartree ensemble approximation revisited: The symmetric phase,''
Phys.\ Rev.\ D {\bf 67}, 116006 (2003)
[arXiv:hep-ph/0208139].

M.~Salle, J.~Smit and J.~C.~Vink,
``Thermalization in a Hartree ensemble approximation to quantum field
dynamics,''
Phys.\ Rev.\ D {\bf 64}, 025016 (2001)
[arXiv:hep-ph/0012346].

  G.~Aarts and J.~Smit,
  ``Particle production and effective thermalization in inhomogeneous mean
  field theory,''
  Phys.\ Rev.\ D {\bf 61}, 025002 (2000)
  [arXiv:hep-ph/9906538].

\bibitem{Baacke:2001zt}
  J.~Baacke and S.~Michalski,
  ``Nonequilibrium evolution in scalar O(N) models with spontaneous  symmetry
  breaking,''
  Phys.\ Rev.\ D {\bf 65}, 065019 (2002)
  [arXiv:hep-ph/0109137].

\bibitem{Calzetta:1986cq}
E.~Calzetta and B.~L.~Hu,
``Nonequilibrium Quantum Fields: Closed Time Path Effective Action, Wigner
Function And Boltzmann Equation,''
Phys.\ Rev.\ D {\bf 37} (1988) 2878.

\bibitem{Cornwall:1974vz}
J.~M.~Cornwall, R.~Jackiw and E.~Tomboulis,
``Effective Action For Composite Operators,''
Phys.\ Rev.\ D {\bf 10}, 2428 (1974).


\bibitem{Baacke:BeyondHF}
J.~Baacke and S.~Michalski,
``O(N) linear sigma model beyond the Hartree approximation at finite
temperature,''
arXiv:hep-ph/0312031.

J.~Baacke and S.~Michalski,
``The O(N) linear sigma model at finite temperature beyond the Hartree
approximation,''
Phys.\ Rev.\ D {\bf 67}, 085006 (2003)
[arXiv:hep-ph/0210060].

J.~Baacke and S.~Michalski, ``Scalar O(N) model at finite temperature:
2PI effective potential in different approximations,''
arXiv:hep-ph/0409153.

J.~Baacke and A.~Heinen,
``Out-of-equilibrium evolution of quantum fields in the hybrid model with
quantum back reaction,''
Phys.\ Rev.\ D {\bf 69}, 083523 (2004)
[arXiv:hep-ph/0311282].

\bibitem{Dominici}
  D.~Dominici and U.~M.~B.~Marconi,
  ``Effective action method for computing next-to-leading corrections of O(N)
  models,''
  Phys.\ Lett.\ B {\bf 319} (1993) 171.

\bibitem{Berges:NLO}
J.~Berges and J.~Serreau,
``Parametric resonance in quantum field theory,''
Phys.\ Rev.\ Lett.\  {\bf 91}, 111601 (2003)
[arXiv:hep-ph/0208070].

J.~Berges and M.~M.~Muller,
``Nonequilibrium quantum fields with large fluctuations,''
arXiv:hep-ph/0209026.

G.~Aarts, D.~Ahrensmeier, R.~Baier, J.~Berges and J.~Serreau,
``Far-from-equilibrium dynamics with broken symmetries from the 2PI-1/N
expansion,''
Phys.\ Rev.\ D {\bf 66}, 045008 (2002)
[arXiv:hep-ph/0201308].

J.~Berges,
``Controlled nonperturbative dynamics of quantum fields out of  equilibrium,''
Nucl.\ Phys.\ A {\bf 699} (2002) 847
[arXiv:hep-ph/0105311].

\bibitem{Cooper:2005vw}
F.~Cooper, J.~F.~Dawson and B.~Mihaila,
``Renormalized broken-symmetry Schwinger-Dyson equations and the 2PI-1/N
expansion for the O(N) model,''
arXiv:hep-ph/0502040.

\bibitem{Lenaghan:1999si}
J.~T.~Lenaghan and D.~H.~Rischke,
``The O(N) model at finite temperature: Renormalization of the gap  equations
in Hartree and large-N approximation,''
J.\ Phys.\ G {\bf 26} (2000) 431
[arXiv:nucl-th/9901049].

\bibitem{Pi:1987df}
S.~Y.~Pi and M.~Samiullah,
``Renormalizability Of The Time Dependent Variational Equations In Quantum
Field Theory,''
Phys.\ Rev.\ D {\bf 36} (1987) 3128.

\bibitem{Cooper:2004rs}
F.~Cooper, B.~Mihaila and J.~F.~Dawson,
``Renormalizing the Schwinger-Dyson equations in the auxiliary field
formulation of lambda phi**4 field theory,''
Phys.\ Rev.\ D {\bf 70}, 105008 (2004)
[arXiv:hep-ph/0407119].

\bibitem{Jakovac}
A.~Jakovac and Z.~Szep,
``Renormalization and resummation in finite temperature field theories,''
arXiv:hep-ph/0405226.

A.~Jakovac and Z.~Szep,
``Renormalization and resummation in field theories,''
arXiv:hep-ph/0408360.

\bibitem{Blaizot}
J.~P.~Blaizot, E.~Iancu and U.~Reinosa,
``Renormalizability of Phi-derivable approximations in scalar phi**4
theory,''
Phys.\ Lett.\ B {\bf 568}, 160 (2003)
[arXiv:hep-ph/0301201].

J.~P.~Blaizot, E.~Iancu and U.~Reinosa,
``Renormalization of phi-derivable approximations in scalar field  theories,''
Nucl.\ Phys.\ A {\bf 736}, 149 (2004)
[arXiv:hep-ph/0312085].

\bibitem{Berges:2005hc}
J.~Berges, S.~Borsanyi, U.~Reinosa and J.~Serreau,
``Nonperturbative renormalization for 2PI effective action
techniques,''
arXiv:hep-ph/0503240.

\bibitem{vanHees}
H.~van Hees and J.~Knoll,
``Renormalization of self-consistent Phi-derivable approximations,''
arXiv:hep-ph/0210262.

\bibitem{Destri:2005qm}
C.~Destri and A.~Sartirana,
``The renormalized and renormalization-group invariant Hartree-Fock
approximation,''
arXiv:hep-ph/0504029.
 
H.~van Hees and J.~Knoll,
``Renormalization of self-consistent Schwinger-Dyson equations at finite
temperature,''
arXiv:hep-ph/0202263.

\bibitem{Baym:1977qb}
G.~Baym and G.~Grinstein,
``Phase Transition In The Sigma Model At Finite Temperature,''
Phys.\ Rev.\ D {\bf 15} (1977) 2897.

\bibitem{vanHees:2002bv}
H.~van Hees and J.~Knoll,
``Renormalization in self-consistent approximation schemes at finite
temperature. III: Global symmetries,''
Phys.\ Rev.\ D {\bf 66} (2002) 025028
[arXiv:hep-ph/0203008].

Y.~B.~Ivanov, F.~Riek and J.~Knoll,
``Gapless Hartree-Fock resummation scheme for the O(N) model,''
arXiv:hep-ph/0502146.

Y.~B.~Ivanov, F.~Riek, H.~van Hees and J.~Knoll,
``Renormalized phi-derivable approximations to theory with spontaneously
broken O(N) symmetry,''
arXiv:hep-ph/0506157.

\bibitem{Eboli:1987fm}
O.~J.~P.~Eboli, R.~Jackiw and S.~Y.~Pi,
``Quantum Fields Out Of Thermal Equilibrium,''
Phys.\ Rev.\ D {\bf 37}, 3557 (1988).

\bibitem{Cooper:1986wv}
F.~Cooper, S.~Y.~Pi and P.~N.~Stancioff,
``Quantum Dynamics In A Time Dependent Variational Approximation,''
Phys.\ Rev.\ D {\bf 34} (1986) 3831.

\bibitem{Cooper:1995zs}
F.~Cooper,
``Nonequilibrium problems in quantum field theory and Schwinger's closed time
path formalism,''
arXiv:hep-th/9504073.

\bibitem{Chou:1984es}
K.~c.~Chou, Z.~b.~Su, B.~l.~Hao and L.~Yu,
``Equilibrium And Nonequilibrium Formalisms Made Unified,''
Phys.\ Rept.\  {\bf 118} (1985) 1.

\bibitem{Baacke:1996se}
J.~Baacke, K.~Heitmann and C.~Patzold,
``Nonequilibrium dynamics: A renormalized computation scheme,''
Phys.\ Rev.\ D {\bf 55} (1997) 2320
[arXiv:hep-th/9608006 ].

\bibitem{Baacke:1997zz}
J.~Baacke, K.~Heitmann and C.~Patzold,
``On the choice of initial states in nonequilibrium dynamics,''
Phys.\ Rev.\ D {\bf 57}, 6398 (1998)
[arXiv:hep-th/9711144].

\end{thebibliography}
\end{document}